\documentclass[11pt,a4paper]{article}

\usepackage{graphics}
\usepackage[pdftex]{graphicx,color}  
\usepackage{fancybox}
\usepackage{amsmath,latexsym,psfrag,epsf,epsfig,amssymb,rotating}
\usepackage{multirow}
\usepackage{url}
\usepackage{natbib}
\usepackage{fullpage}

\newcommand{\pref}[1]{%
    \ref{#1} \ifnum\count0=\pageref{#1}\relax%
    \else (page \pageref{#1})\fi}

\newcommand{\eref}[1]{%
        \ref{#1}\ifnum\count0=\pageref{#1}\relax%
        \else {, p.\pageref{#1}}\fi}

\newcommand{\comment}[1]{}

\newenvironment{algorithm}{\vspace{5 mm}\sc}{\vspace{5 mm}}
\newlength{\labwidth}
\newcommand{\step}[1]{%
    \settowidth{\labwidth}{#1\ }%
    \par\noindent%
    \global\hangindent\labwidth {#1}%
    \hbox{ }%
    }%

\newcommand{\bftheta}{\hbox{\boldmath$\theta$}}
\newcommand{\bfepsilon}{\hbox{\boldmath$\epsilon$}}

\newcommand{\bfI}{\hbox{\boldmath$I$}}

\newcommand{\bfy}{\hbox{\boldmath$y$}}
\newcommand{\bfY}{\hbox{\boldmath$Y$}}

\newcommand{\EE}{{\mathbb{E}}}

\begin{document}

\title{\Large Bayesian inference for exponential random graph models}
\author{Alberto Caimo \& Nial Friel\\
       School of Mathematical Sciences,\\
       University College Dublin, Ireland\\
       \normalsize\texttt{\{alberto.caimo,nial.friel\}@ucd.ie}}
\date{\today}
\maketitle

\begin{center}
 {\Large Bayesian inference for exponential random graph models}\vspace*{1cm}
\end{center}

\begin{abstract} 

Exponential random graph models are extremely difficult models to handle from a 
statistical viewpoint, since their normalising constant, which depends on model 
parameters, is available only in very trivial cases. We show how inference can 
be carried out in a Bayesian framework using a MCMC algorithm, which circumvents 
the need to calculate the normalising constants. We use a population MCMC approach 
which accelerates convergence and improves mixing of the Markov chain. 
This approach improves performance with respect to the Monte Carlo maximum 
likelihood method of \cite{gey:tho92}. 

\end{abstract}

\section{Introduction}

Our motivation for this article is to propose Bayesian inference for estimating
exponential random graph models (ERGMs) which are some of the most important models 
in many research areas such as social networks analysis, physics and biology.
The implementation of estimation methods for these models is a key point which 
enables us to use parameter estimates as a basis for simulation and to reproduce 
features of real networks.
Unfortunately, the intractability of the normalising constant and degeneracy are two 
strong barriers to parameter estimation for these models. 
The classical inferential methods such as the Monte Carlo maximum likelihood (MC-MLE)
of \cite{gey:tho92} and pseudolikelihood estimation (MPLE) of \cite{bes74} and 
\cite{str:ike90}, are very widely used in partice, but are lacking in certain 
respects. In some instance it is difficult to obtain reasonable results or to 
understand the properties of the approximations, using these methods.

The Bayesian estimation approach for these models has not yet been deeply and fully 
explored. Despite this, Bayesian inference is very appropriate in this context since 
it allows uncertainty about model parameters given the data to be explored through a 
posterior distribution. Moreover the Bayesian approach allows a formal comparison 
procedure among different competing models using posterior probabilities. Therefore 
the methods presented in this work aim to contribute to the development of a 
Bayesian-based methodology area for these models.

Networks are relational data represented as graphs, consisting of nodes and edges.
Many probability models have been proposed in order to summarise the general structure 
of graphs by utilising their local properties. 
Each of these models take different assumptions into account: the Bernoulli random 
graph model \citep{erd:ren59} in which edges are considered independent of each other; 
the $p_1$ model \citep{hol:lei81} where dyads are assumed independent, and its random 
effects variant the $p_2$ model \citep{van:sni:zij04}; and the Markov random graph 
model \citep{fra:str86} where each pair of edges is conditionally dependent given the 
rest of the graph. The family of exponential random graph models (\cite{was:pat96}, 
see also \cite{rob:sni:wan:han:pat07} for a recent review) is a generalisation of the 
latter model and has been thought to be a flexible way to model the complex dependence 
structure of network graphs. Recent developments have led to the introduction of new 
specifications by \cite{sni:pat:rob:han06} and the implementation of curved exponential 
random graph models by \cite{hun:han06}. New modelling alternatives such as the latent 
variable models have been proposed by \cite{hof:raf:han02} under the assumption that 
each node of the graph has a unknown position in a latent space and the probability of 
the edges are functions of those positions and node covariates.
The latent position cluster model of \cite{han:raf:tan07} represents a further extension 
of this approach that takes account of clustering.
Stochastic blockmodel methods \citep{now:sni01} involve block model structures whereby
nodes of the graph are partitioned into latent classes and their relationship depends 
on block membership. More recently, the mixed membership approach \citep{air:ble:fie:xin08} 
has emerged as a flexible modelling tool for networks extending the assumption of a single 
block membership.

In this paper we are concerned with Bayesian inference for exponential random graph models. 
This leads to the problem of the double intractability of the posterior density as both 
the model and the posterior normalisation terms cannot be evaluated. 
This problem can be overcome by adapting the exchange algorithm of \cite{Murray06} to the 
network graph framework. Typically the high posterior density region is thin and highly 
correlated. For this reason, we also propose to use a population-based MCMC approach so as 
to improve the mixing and local moves on the high posterior density region reducing the 
chain's autocorrelation significantly. 
An \texttt{R} package called \texttt{Bergm}, which accompanies this paper, 
contains all the procedures used in this paper. It can be found at: 
\texttt{http://cran.r-project.org/web/packages/Bergm/}.

This article is structured as follows. A basic description of exponential random graph models 
together with a brief illustration of the issue of degeneracy is given in Section 2. 
Section 3 reviews two of the most important and used methods for likelihood inference. 
In Section 4 we introduce Bayesian inference via the exchange algorithm and its further 
improvement through a population-based MCMC procedure. 
In Section 5 we fit different models to three benchmark network datasets and outline some 
conclusions in Section 6. 

\section{Exponential random graph models}

Typically networks consist of a set of actors and relationships between pairs of them, 
for example social interactions between individuals. 
The network topology structure is measured by
a random adjacency matrix $\bfY$ of a graph on $n$ nodes (actors) and a set of edges 
(relationships) $\{ Y_{ij}: i=1,\dots,n; j=1,\dots,n\}$ where $Y_{ij}=1$ if the 
pair $(i,j)$ is connected, and $Y_{ij}=0$ 
otherwise. Edges connecting a node to itself are not allowed so $Y_{ii}=0$. 
The graph $\bfY$ may be directed (digraph) or undirected depending on the 
nature of the relationships between the actors.\\
Let $\mathcal{Y}$ denote the set of all possible graphs on $n$ nodes and let $\bfy$ a 
realisation of $\bfY$. Exponential random graph models (ERGMs) are a particular class 
of discrete linear exponential families which represent the probability distribution 
of $\bfY$ as
\begin{equation}
\pi(\bfy|\bftheta) = \frac{q_{\bftheta}(\bfy)}{z(\bftheta)} = 
\frac{\exp\{\bftheta^t s(\bfy)\}}{z(\bftheta)}
\label{eqn:ergm}
\end{equation}
where $s(\bfy)$ is a known vector of sufficient statistics (e.g. the number of edges, 
degree statistics, triangles, etc.) and $\bftheta \in \Theta$ are model parameters.
Since $\mathcal{Y}$ consists of $2^{\binom{n}{2}}$ possible undirected graphs, the 
normalising constant $z(\bftheta) = \sum_{\bfy \in \mathcal{Y}} \exp\{\bftheta^t s(\bfy)\}$ 
is extremely difficult to evaluate for all but trivially small graphs. For this reason, 
ERGMs are very difficult to handle in practice. In spite of this difficulty, ERGMs are 
very popular in the literature since they are conceived to capture the complex dependence 
structure of the graph and allow a reasonable interpretation of the observed data.
The dependence hypothesis at the basis of these models is that edges self organize into 
small structures called configurations. 
There is a wide range of possible network configurations \citep{rob:pat:kal:lus07} which 
give flexibility to adapt to different contexts.
A positive parameter value for $\theta_i \in \bftheta$ result in a tendency for the 
certain configuration corresponding to $s_i(\bfy)$ to be observed in the data than would 
otherwise be expected by chance.

\subsection{Degeneracy}

Degeneracy is one of the most important aspects of random graph models: it 
refers to a probability model defined by a certain value 
of $\bftheta$ that places most of its mass on a small number of graph topologies, 
for example, empty graphs or complete graphs.

Consider the mean value parametrisation for the model (\ref{eqn:ergm}) defined by 
$\mu(\bftheta) = \EE_{\bftheta}[s(\bfy)]$.
Let $C$ be the convex hull of the set $\{s(\bfy):\bfy \in \mathcal{Y}\}$, $ri(C)$ its
relative interior and $rbd(C)$ its relative boundary. 
It turns out that if the expected values $\mu(\bftheta)$ of the sufficient statistics 
$\mu(\bftheta)$ approach the boundary $rbd(C)$ of the hull, the model 
places most of the probability mass on a small set of graphs belonging to 
$deg(\mathcal{Y}) = \{ \bfy \in \mathcal{Y}:s(\bfy) \in rbd(C) \}$. 
It is also known that the MLE exists if and only if $s(\bfy) \in ri(C)$ and if it exists 
it is unique.

When the model is near-degenerate, MCMC inference methods may fail to find the maximum 
likelihood estimate (MLE) and returning an estimate of $\bftheta$ that is unlikely to 
generate networks closely resembling the observed graph.
This is because the convergence of the algorithm may be greatly affected by degenerate 
parameters values which during the network simulation process may yield graphs which are 
full or empty.

A more detailed description of this issue can be found in \cite{han03} and \cite{rin:fie:zho09}.
The new specifications proposed by \cite{sni:pat:rob:han06} can often mitigate degeneracy 
and provide reasonable fit to the data.

\section{Classical inference}

\subsection{Maximum pseudolikelihood estimation}
A standard approach to approximate the distribution of a Markov random field is to 
use a maximum pseudolikelihood (MPLE) approximation, first proposed in \cite{bes74} 
and adapted for social network models in \cite{str:ike90}.
This approximation consists of a product of easily normalised full-conditional 
distributions
\begin{align}
\pi(\bfy|\bftheta) \approx \pi_{pseudo}(\bfy|\bftheta) 
&=\prod_{i\neq j} \pi(y_{ij}|\bfy_{-ij},\bftheta) 
\label{eqn:pseudo}
\\
&=\prod_{i\neq j}
\frac{\pi(y_{ij}=1|\bfy_{-ij},\bftheta)^{y_{ij}}}
{\left[ 1-\pi(y_{ij}=0|\bfy_{-ij},\bftheta)\right]^{y_{ij}-1}}\notag
\end{align}
where $\bfy_{-ij}$ denotes all the dyads of the graph excluding $y_{ij}$.
The basic idea underlying this method is the assumption of weak dependence between 
the variables in the graph so that the likelihood can be well approximated by the 
pseudolikelihood function. This leads to fast estimation.  Nevertheless this 
approach turns out to be generally inadequate since it only uses local information 
whereas the structure of the graph is affected by global interaction. In the context of
the autologistic distribution in spatial statistics, \cite{fri:pet:rev09} showed that 
the pseudolikelihood estimator can lead to inefficient estimation.

\subsection{Monte Carlo maximum likelihood estimation}

There are many instances of statistical models with intractable normalising constants. 
This general problem was tackled in \cite{gey:tho92} who introduced the Monte Carlo 
maximum likelihood (MC-MLE) algorithm. In the context of ERGMs, a key identity is the 
following
\begin{equation}
\frac {z(\bftheta)} {z(\bftheta_0)} 
= \EE_{\bfy|\bftheta_0} 
\left[ 
\frac {q_{\bftheta}(\bfy)} {q_{\bftheta_0}(\bfy)} 
\right] 
= \sum_{\bfy} 
\frac {q_{\bftheta}(\bfy)} {q_{\bftheta_0}(\bfy)} 
\frac {q_{\bftheta_0}(\bfy)} {z(\bftheta_0)} \label{eqn:key}
\end{equation}
where $q_{\bftheta}(\cdot)$ is the unnormalised likelihood of parameters $\bftheta$, $\bftheta_0$ is fixed set of parameter 
values, and $\EE_{\bfy|\bftheta_0}$ denotes an expectation taken with respect to the 
distribution $\pi(\bfy|\bftheta_0)$. In practice this ratio of normalising constants is 
approximated using graphs $\bfy_1,\dots,\bfy_m$  sampled via MCMC from $\pi(\bfy|\bftheta_0)$ 
and importance sampling. This yields the following approximated log likelihood ratio:
\begin{equation} 
w_{\bftheta_0}(\bftheta)=\ell(\bftheta)-\ell(\bftheta_0) \approx (\bftheta - \bftheta_0)^t s(\bfy) - 
\log\left[ \frac{1}{m}\sum_{i=1}^m \exp\left\{ (\bftheta-\bftheta_0)^t s(\bfy_i) \right\} \right] 
\label{eqn:gey:thom}
\end{equation}
where $\ell(\cdot)$ is the log-likelihood. $w_{\bftheta_0}$ is a function of $\bftheta$, and 
its maximum value serves as a Monte Carlo estimate of the MLE.

A crucial aspect of this algorithm is the choice of $\bftheta_0$. Ideally $\bftheta_0$ should be 
very close to the maximum likelihood estimator of $\bftheta$. Viewed as a function of $\bftheta$, 
$w_{\bftheta_0}(\bftheta)$ in (\ref{eqn:gey:thom}) is very sensitive to the choice of $\bftheta_0$. 
A poorly chosen value of $\bftheta_0$ may lead to an objective function (\ref{eqn:gey:thom}) that 
cannot even be maximised. We illustrate this point in Section 3.4. 

In pratice, $\bftheta_0$ is often chosen as the maximiser of (\ref{eqn:pseudo}), although this 
itself maybe a very biased estimator. Indeed, (\ref{eqn:gey:thom}) may also be sensitive to 
numerical instability, since it effectively computes the ratio of a normalising constant, but 
it is well understood that the normalising constants can vary by orders of magnitude with $\bftheta$.

In fact, if the value of $\bftheta_0$ lies in the ``degenerate region'' then graphs 
$\bfy_i,\dots,\bfy_m$ simulated from $\pi(\bfy|\bftheta_0)$ will tend to belong to $deg(\mathcal{Y})$
and hence the estimate of the ratio of normalising constants $z(\bftheta)/z(\bftheta_0)$ will be
very poor. Consequently $w_{\bftheta_0}(\bftheta)$ in (\ref{eqn:gey:thom}) will yield unreasonable 
estimates of the MLE. In the worst situation, $w_{\bftheta_0}(\bftheta)$ does not have an optimum. 
This behaviour is well understood as presented in \cite{han03} and \cite{rin:fie:zho09}.

\subsection{A pedagogical example}
\begin{figure}[htp]
\centering
\includegraphics[scale=0.6]{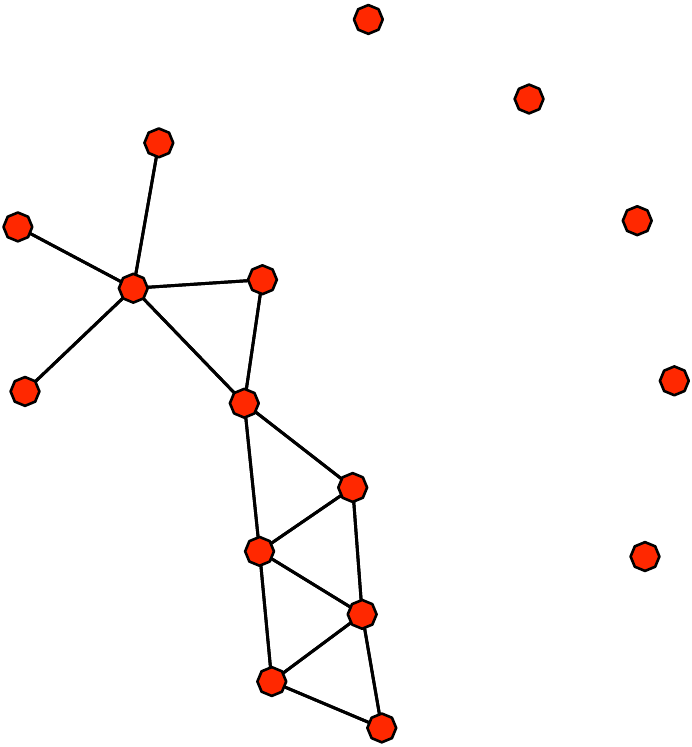}
\caption{Florentine family business graph.}
\label{fig:graph-flo}
\end{figure}

Let us consider, for the purpose of illustration, a simple 16-node graph: the Florentine family 
business graph \citep{pad:ans93} concerning the business relations between some Florentine families 
in around 1430. The network is displayed in Figure~\ref{fig:graph-flo} and shows 
that few edges between the families are present but with quite high level of triangularisation.


\begin{figure}[htp]
\centering
\includegraphics[scale=0.75]{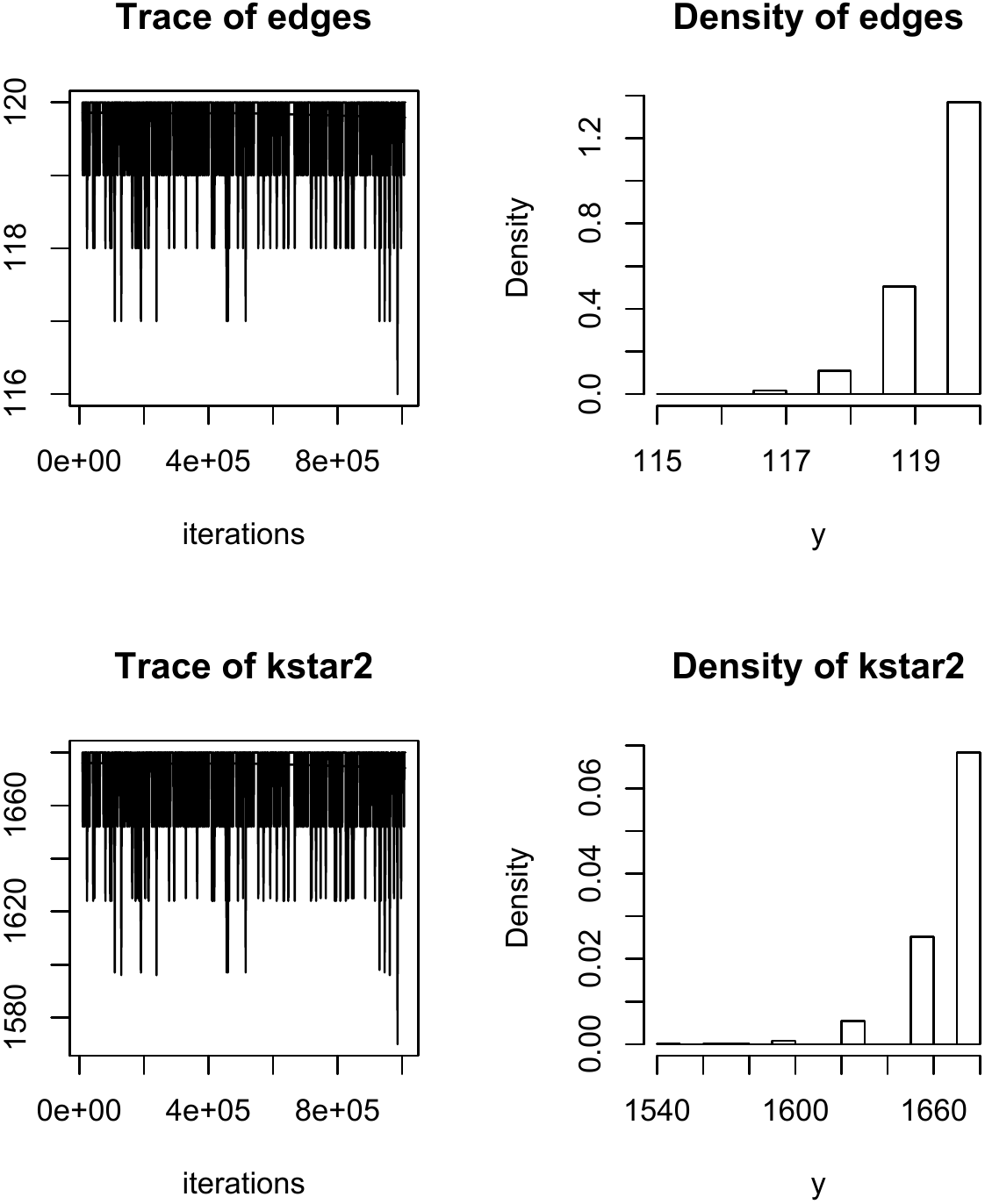}
\caption{MCMC output of the graphs simulated from (\ref{eqn:two-star}) where $\bftheta_0$ is
the maximum pseudolikelihood estimate. The top and bottom rows corresponds to $s_1(\bfy)$ and
$s_2(\bfy)$, respectively.}
\label{fig:mcmle-flo}
\end{figure}

For this undirected network we propose to estimate the following 2-dimensional model:
\begin{equation}
\pi(\bfy|\bftheta) = \frac{1}{z(\bftheta)} \exp \left\lbrace  \theta_1 s_1(\bfy) + \theta_2 s_2(\bfy) 
\right\rbrace 
\label{eqn:two-star}
\end{equation}
with statistics $s_1(\bfy) = \sum_{i<j}y_{ij}$ and $s_2(\bfy) = \sum_{i<j<k}y_{ik}y_{jk}$ which
are respectively the observed number of edges and two-stars, that is, the number of nodes with
degree at least two.
We use the \texttt{statnet} package \citep{han:hun:but:goo:mor07} in the \texttt{R} programming language
to compute estimates of the maximum pseudolikelihood estimator and the Markov chain Monte Carlo estimator.
Despite its simplicity the parameters of this model turn out to be difficult to estimate in practice, 
and both the MPLE and the MC-MLE fail to produce reasonable estimates.
Table~\ref{tab:tabmcmle-flo} reports the parameter estimates of both MPLE and MC-MLE.
The standard errors of the MC-MLE are unreasonably large due to poor convergence of the algorithm. 
In fact, in this case it turns out that graphs $\bfy_1,\dots,\bfy_m$ simulated from 
$\pi(\bfy|\bftheta_0)$, which are used in (\ref{eqn:gey:thom}), are typically complete graphs.
This is illustrated in Figure~\ref{fig:mcmle-flo}. 
The left-hand plots show the trace of the MCMC sampled sufficient statistics simulated during 
the MCMC iterations, where $\bftheta_0$ is set equal to the MPLE. The right-hand plots show the respective 
density estimate. These parameter estimates generate mainly full graphs. This is good evidence of degeneracy,
thereby illustrating, in this example, that MPLE and MC-MLE do not give parameter estimates which are
in agreement with the observed data. 

\begin{table}[htp]
\centering
\begin{tabular}{c|cc|cc}
 \multicolumn{1}{c}{} & \multicolumn{2}{c}{MC-MLE} & \multicolumn{2}{c}{MPLE}\\
\hline \hline
Parameter & Estimate & Std. Error & Estimate & Std. Error \\ \hline
$\theta_1$ (edges) & -3.39 & 21.69 & -3.39 & 0.70\\
$\theta_2$ (2-stars) & 0.30 & 0.79 & 0.35 & 0.14\\
\hline \hline
\end{tabular}
\caption{Summary of parameter estimates of the model (\ref{eqn:two-star}).}
\label{tab:tabmcmle-flo}
\end{table}

\begin{figure}[htp]
\centering
\includegraphics[scale=0.75]{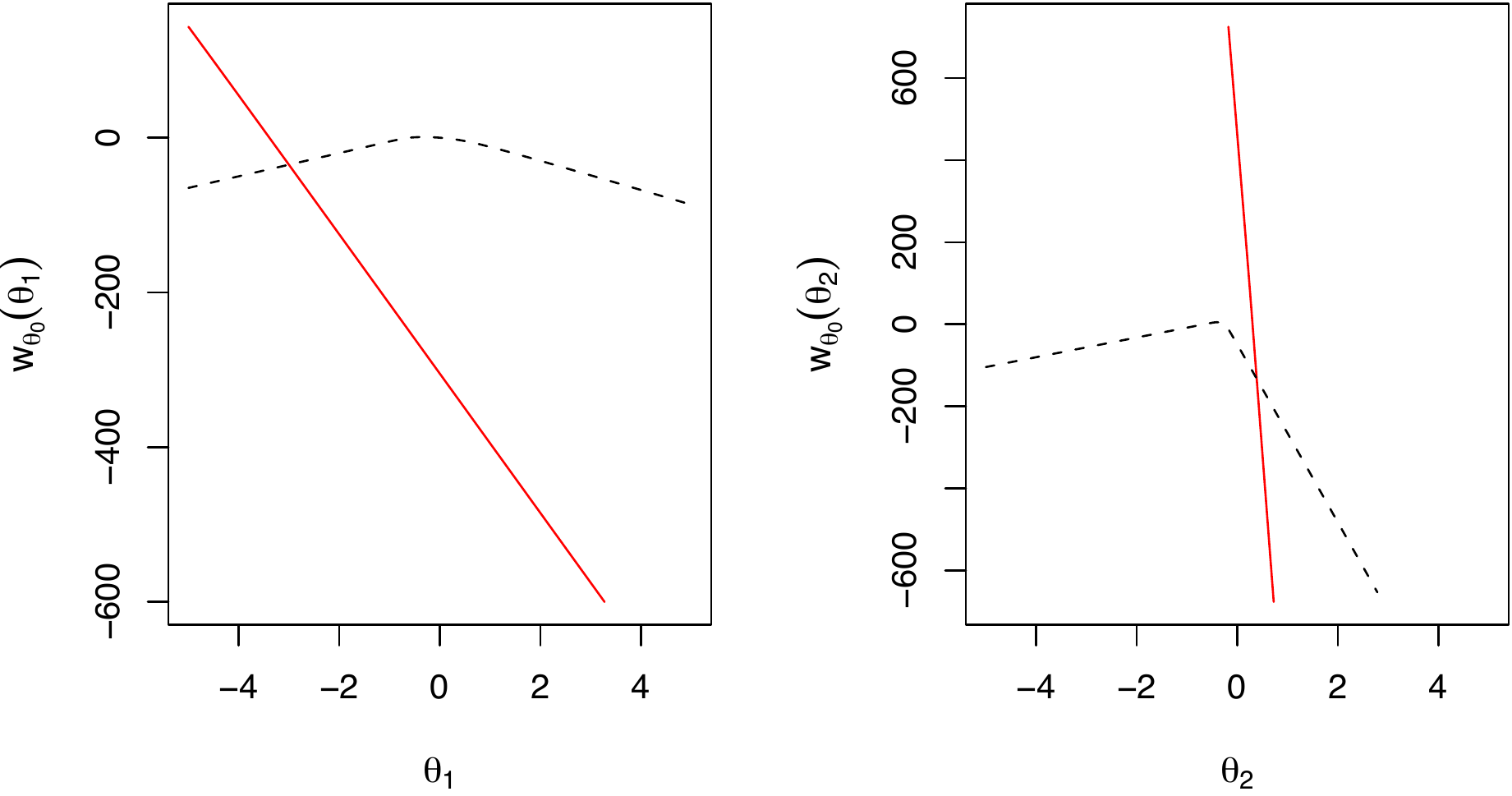}
\caption{Approximation of the log ratio in (\ref{eqn:gey:thom}) 
         by using $w_{\bftheta_0}(\bftheta)$ for two different initial values: 
         $\bftheta_0=\;\mbox{MPLE}$ (solid curve) and 
         $\bftheta_0=(0,0)$ (dotted curve).}
\label{fig:mcmlefail-flo}
\end{figure}

As before, the issue of choice of initial value $\bftheta_0$ is crucial to the performance 
and convergence of MC-MLE. 
In Figure~\ref{fig:mcmlefail-flo} it can be observed how $w_{\bftheta_0}(\bftheta)$ in (\ref{eqn:gey:thom}) 
varies with respect to two possible choices of $\bftheta_0$. 
In this case note that when $\bftheta_0$ corresponds to the MPLE, (\ref{eqn:gey:thom}) gives an 
approximation that cannot even be maximized whereas $\bftheta_0=(0,0)$, for example, would 
seem to be a better choice.
 
\section{Bayesian inference}

Consider the Bayesian treatment of this problem (see \cite{kos04}), 
where a prior distribution $\pi(\bftheta)$ is placed
on $\bftheta$, and interest is in the posterior distribution
\begin{equation*}
\pi(\bftheta | \bfy ) \propto \pi(\bfy|\bftheta) \pi(\bftheta).
\end{equation*}
This type of posterior distribution is sometimes called ``doubly-intractable'' due to the 
(standard) intractability of sampling directly from the posterior distribution, but also due
to the intractability of the likelihood model within the posterior. 

A na\"{i}ve implementation of a Metropolis-Hastings algorithm proposing to move from $\bftheta$ to
$\bftheta'$ would require calculation of the following ratio at each sweep of the algorithm
\begin{equation} 
\frac{q_{\bftheta'}(\bfy)\pi(\bftheta')}{q_{\bftheta}(\bfy)\pi(\bftheta)} \times \frac{z(\bftheta)}{z(\bftheta')} 
\label{eqn:naive}
\end{equation}
which is unworkable due to the presence of the normalising constants $z(\bftheta)$ and $z(\bftheta')$.

\subsection{The exchange algorithm}
There has been considerable recent activity on the problem of sampling from such complicated
distributions, for example, \cite{mol:pet06}. The algorithm presented in this paper overcomes the
problem of sampling from a distribution with intractable normalising constant, to a large extent. 
However the algorithm can result in an MCMC chain with poor mixing among the parameters. 
In social network analysis, further developments of this approach have been proposed in \cite{kos08} 
and \cite{kos:rob:pat09} by introducing the linked importance sampler auxiliary variable (LISA) algorithm 
which employs an importance sampler in each iteration to estimate the acceptance probability.

In this paper we adapt to the ERGMs context the simple and easy-to-implement exchange algorithm presented 
in \cite{Murray06}. The algorithm samples from an augmented distribution
\begin{equation}
\pi(\bftheta',\bfy',\bftheta|\bfy) \propto \pi(\bfy|\bftheta)\pi(\bftheta) h(\bftheta'|\bftheta) \pi(\bfy'|\bftheta') 
\label{eqn:exchange}
\end{equation}
where $\pi(\bfy'|\bftheta')$ is the same distribution as the original distribution on which the data $\bfy$
is defined. The distribution $h(\bftheta'|\bftheta)$ is any arbitrary distribution for the augmented variables 
$\bftheta'$ which might depend on the variables $\bftheta$, for example, a random walk distribution centered 
at $\bftheta$. It is clear that the marginal distribution for $\bftheta$ in (\ref{eqn:exchange}) is the posterior 
distribution of interest. The algorithm can be written in the following concise way:

\begin{algorithm}
\step{1.}Gibbs update of $(\bftheta',\bfy')$:\\
\textit{i} Draw $\bftheta' \sim h(\cdot|\bftheta)$.\\
\textit{ii} Draw $\bfy' \sim \pi(\cdot|\bftheta')$.\\
\step{2.}Propose the exchange move from $\bftheta$ to $\bftheta'$ with probability
\begin{equation}
\alpha=\min\left( 1, \frac{q_{\bftheta}(\bfy')\pi(\bftheta') h(\bftheta|\bftheta') q_{\bftheta'}(\bfy)}
{q_{\bftheta}(\bfy)\pi(\bftheta) h(\bftheta'|\bftheta) q_{\bftheta'}(\bfy')} 
\times \frac{z(\bftheta)z(\bftheta')}{z(\bftheta)z(\bftheta')} \right).
\label{eqn:alpha}
\end{equation}
\end{algorithm}

Notice in step 2, that all intractable normalising constants cancel above and below the fraction.
In practice, the exchange move proposes to offer the observed data $\bfy$ the auxiliary parameter $\bftheta'$
and similarly to offer the auxiliary data $\bfy'$ the parameter $\bftheta$.
The affinity between $\bftheta$ and $\bfy'$ is measured by $q_{\bftheta}(\bfy')/q_{\bftheta}(\bfy)$ and the
affinity between $\bftheta'$ and $\bfy$ by $q_{\bftheta'}(\bfy)/q_{\bftheta'}(\bfy')$.
The difficult step of the algorithm is 1.\textit{ii} since this requires a draw from $\pi(\bfy'|\bftheta')$.
Note that perfect sampling (\cite{pro:wil96}) is often possible for Markov random field models, however a 
pragmatic alternative is to sample from $\pi(\bfy'|\bftheta')$ by standard MCMC methods, for example, 
Gibbs sampling, and take a realisation from a long run of the chain as an approximate draw from the distribution.
If we suppose that the proposal density $h(\cdot)$ is symmetric, then we can write the acceptance ratio $\alpha$ 
in (\ref{eqn:alpha}) as
\begin{equation*}
\alpha=\min\left( 1, \frac{q_{\bftheta'}(\bfy)\pi(\bftheta')q_{\bftheta}(\bfy')}
{q_{\bftheta}(\bfy)\pi(\bftheta)q_{\bftheta'}(\bfy')} \right).
\end{equation*}
Comparing this to (\ref{eqn:naive}), we see that $q_{\bftheta}(\bfy')/q_{\bftheta'}(\bfy')$ can be thought of 
as an importance sampling estimate of the ratio $z(\bftheta)/z(\bftheta')$ since
\begin{equation*} 
\EE_{\bfy'|\bftheta'}\frac{q_{\bftheta}(\bfy')}{q_{\bftheta'}(\bfy')} = 
\sum_{\bfy} 
\frac{q_{\bftheta}(\bfy')}{q_{\bftheta'}(\bfy')} 
\frac{q_{\bftheta'}(\bfy')}{z(\bftheta')}
= \frac{z(\bftheta)}{z(\bftheta')}.
\end{equation*}
Recall that this is the identity (\ref{eqn:key}) used in the Geyer-Thompson method.
Note that this algorithm has some similarities with another likelihood-free method called 
Approximate Bayesian Computation (ABC) (see \cite{bea:zha:bal02} and \cite{mar:mol:pla:tav03}). 
The ABC algorithm also relies on drawing parameter 
values $\bftheta'$ and simulating new graphs from those values. 
The proposed move is accepted if there is good agreement
between auxiliary data and observed data in terms of
summary statistics. Here a good approximation to the true posterior density is guaranteed by
the fact that the summary statistics are sufficient statistics of
the probability model.

\subsection{MCMC simulation from ERGMs}
Recall that step 1.\textit{ii} of the exchange algorithm requires a draw $\bfy'$ from $\pi(\bfy'|\bftheta)$. 
The simulation of a network given a parameter value is accomplished by an MCMC algorithm which 
at each iteration compares the probability of a proposed graph to the observed one and then decides
whether or not accept the proposed network. The latter is selected at each step by proposing a change 
in the current state of a single dyad (i.e. creating a new edge or dropping an old edge).
This is obviously a computationally intensive procedure.\\
In order to improve mixing of the Markov chain the \texttt{ergm} package for \texttt{R} 
\citep{hun:han:but:goo:mor08} uses by default the ``tie no tie" (TNT) sampler. This is the approach we take
throughout this paper. At each iteration of the chain the TNT sampler, instead of selecting a dyad at random, 
first selects with equal probability the set of edges or the set of empty dyads, and than swaps a dyad at 
random within that chosen set. In this way, since most of the realistic networks are quite sparse, 
the probability of selecting an empty dyad to swap is lower and the sampler 
does not waste too much time proposing new edges which are likely to be rejected \citep{mor:han:hun08}.
Moreover, initialising the auxiliary chain with the observed network $\bfy$ leads to improved efficiency
in terms of acceptance rates. This is the default choice in our implementation of the algorithm, which we
now describe.

\subsection{Implementing the algorithm}
The exchange algorithm can be easily implemented by using existing software and the 
\texttt{ergm} package is particularly appropriate for this purpose.
An \texttt{R} package called \texttt{Bergm}\footnote{\texttt{http://cran.r-project.org/web/packages/Bergm/}}, 
which accompanies this paper, draws heavily on the \texttt{ergm} package and provides functions for the 
implementation of the exchange algorithm which have been used to obtain the results reported in this paper.

\begin{figure}[htp]
\centering
\includegraphics[scale=0.7]{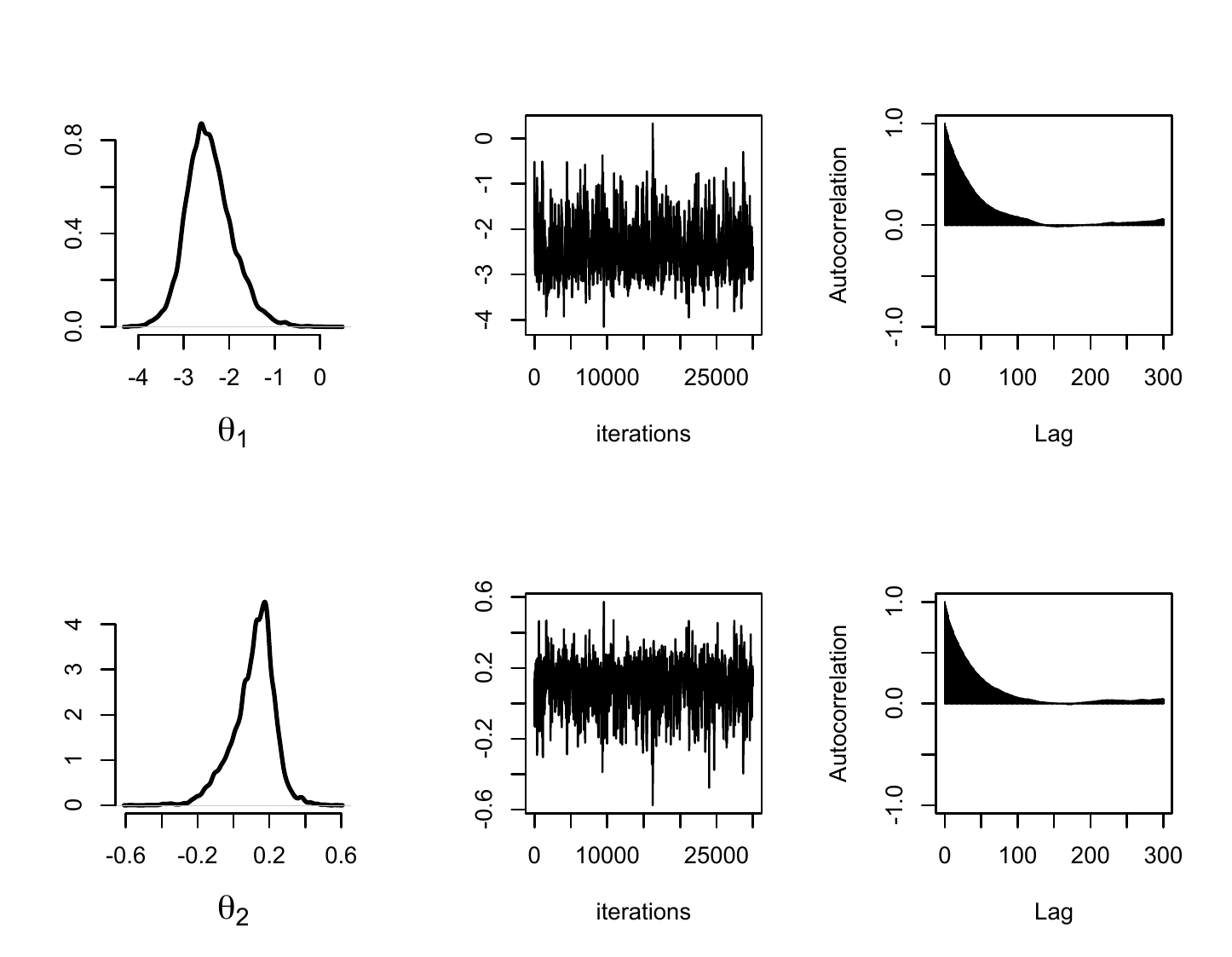}
\caption{MCMC output of the pedagogical model (\ref{eqn:bay-two-star}).}
\label{fig:ssumcmc-flo}
\end{figure}

Let us return to the running example from Section 3.4.
The posterior distribution of model (\ref{eqn:two-star}) is written as	
\begin{equation}
\pi(\bftheta|\bfy) 
\propto 
\frac{1}{z(\bftheta)} 
\exp 
\left\lbrace  \theta_1 \sum_{i<j}y_{ij} + \theta_2 \sum_{i<j<k}y_{ik}y_{jk} 
\right\rbrace 
\pi(\bftheta).
\label{eqn:bay-two-star}
\end{equation}
Here we assume a flat multivariate normal prior $\pi(\bftheta)\sim\mathcal{N}(\mathbf{0},30\bfI_d)$ 
where $\bfI_d$ is the identity matrix of size equal to the number of model dimensions $d$.
The simplest approach to update the parameter state is to update its components one at a time,
using a single-site update sampler.
We use the following proposals for the first and the second parameter value, respectively,
$h( \cdot | \theta_1 ) \sim \mathcal{N}(0,1)$ and $h( \cdot | \theta_2 ) \sim \mathcal{N}(0,0.1)$
where the proposal tuning parameters are chosen in order to reach a reasonable level of 
mixing, in this case an overall acceptance rate of $18\%$.
The auxiliary chain consists of $1,000$ iterations and the main one of $30,000$ iterations.\\
The posterior estimates are reported in Table~\ref{tab:tabaux-flo}.
From Figure~\ref{fig:ssumcmc-flo} it can be seen that the estimates produced by the exchange 
algorithm are very different from the ones obtained by MC-MLE and MPLE.
The autocorrelation plots show that the autocorrelations are negligible after lag 150.

As mentioned in Section 4.1, an important issue for implementation of this algorithm is that of drawing 
from $\pi(\bfy'|\bftheta')$ and a pragmatic way to do that is to use the final realisation of a graph 
simulated from a long run of an MCMC sampler. Here we are interested in testing the sensitivity of the 
posterior output to the number of auxiliary iterations used to generate the sampled graph. 
Table~\ref{tab:tabaux-flo} displays the results using three different number of iterations for
the auxiliary chain. As we can see from the three outputs, there is not a significant difference 
between the results obtained using $500$ or more iterations.

\begin{table}[htp]
\centering
\begin{tabular}{c|c|cc}
\multicolumn{4}{c}{} \\
\hline \hline
Iterations & Parameter & Post. Mean & Post. Sd.\\ \hline 
\multirow{2}{*}{$500$}
& $\theta_1$ (edges) & -2.43 & 0.52 \\
& $\theta_2$ (2-stars) & 0.11 & 0.12 \\ \hline
\multirow{2}{*}{$1,000$}
& $\theta_1$ (edges) & -2.42 & 0.51 \\
& $\theta_2$ (2-stars) & 0.11 & 0.11 \\ \hline
\multirow{2}{*}{$5,000$}
& $\theta_1$ (edges) & -2.43 & 0.51 \\
& $\theta_2$ (2-stars) & 0.10 & 0.12 \\ \hline \hline
\end{tabular}
\caption{Posterior outputs for different numbers of iterations for the auxiliary chain.}
\label{tab:tabaux-flo}
\end{table}

It is worth mentioning here that the posterior density is highly correlated and that the 
high posterior density region is thin. The high posterior density region for the pedagogical 
example is shown in the left-hand plot of Figure~\ref{fig:2d-flo}.
In such a situation, which is not rare for these kind of models, in absence of any information 
about the covariance structure of the posterior distribution, the single-site update may not 
be the best procedure as it may result in a poorly mixing MCMC sampler. We will discuss this 
issue in Section 4.5.

\subsection{Convergence of the Markov chain}
\begin{figure}[htp]
\centering
\includegraphics[scale=0.55]{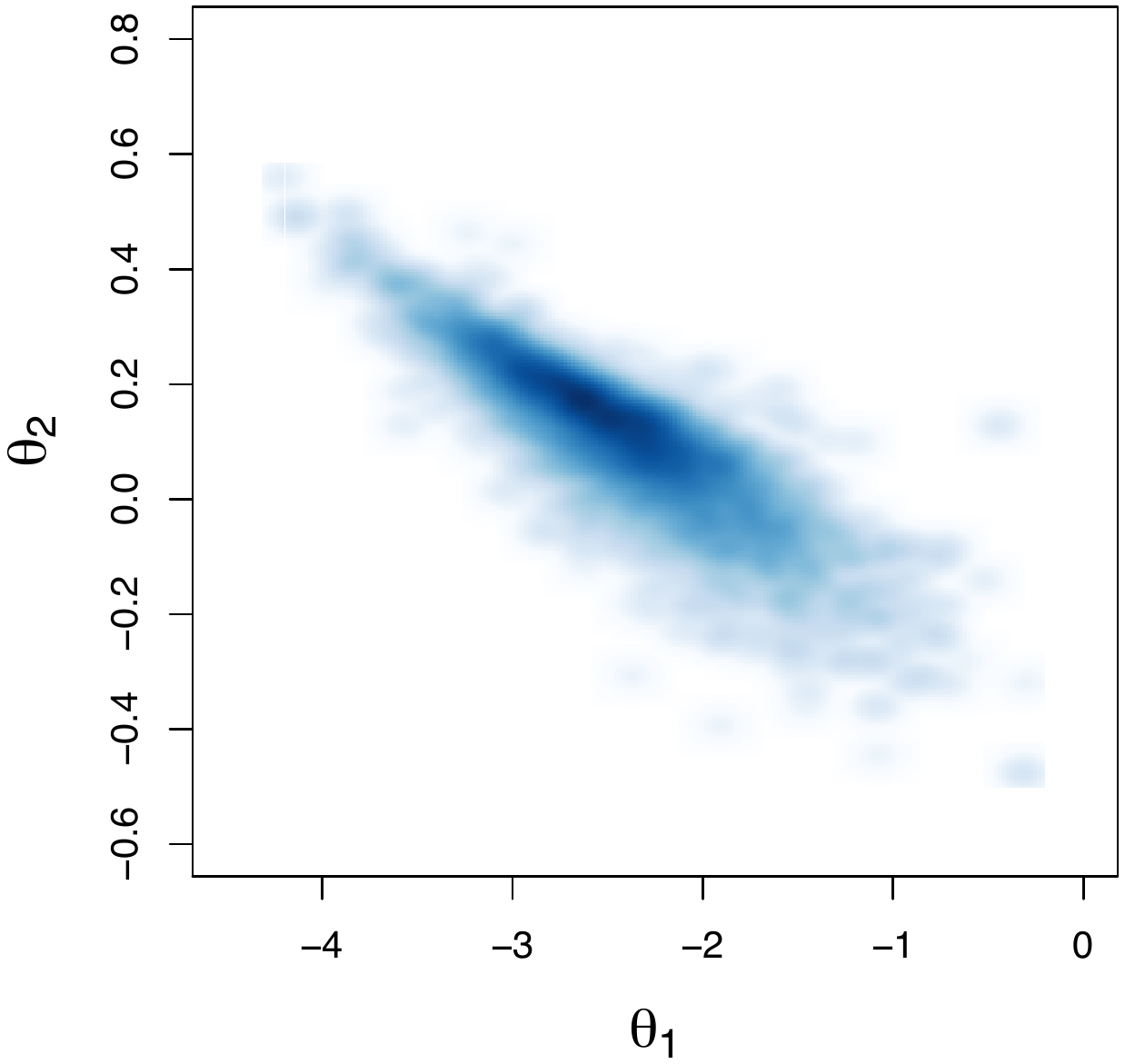}
\hspace{0.7cm}
\includegraphics[scale=0.55]{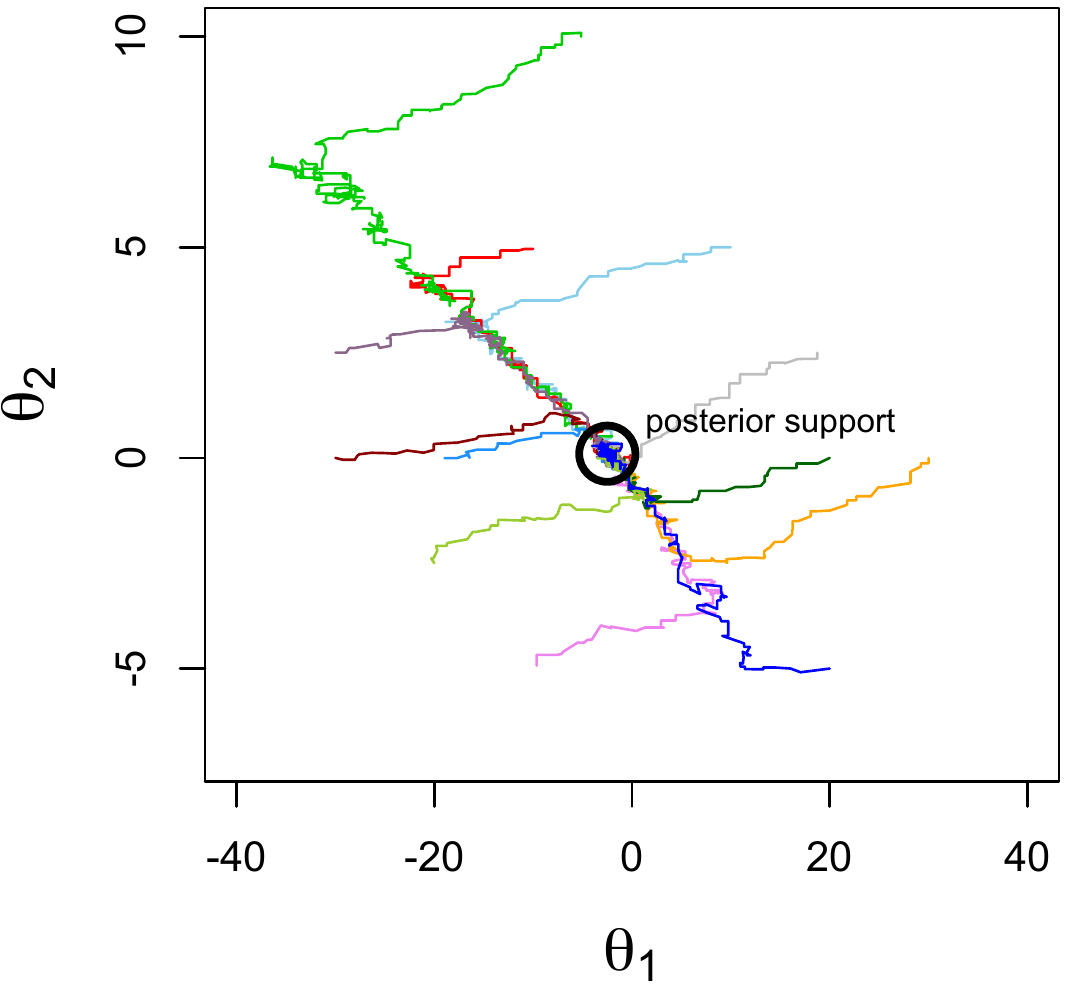}
\caption{Posterior density (left) and traces (right) of the first $1,000$ iterations of 12 chains from 
the posterior distribution corresponding to the pedagogical example.}
\label{fig:2d-flo}
\end{figure}

Exploration of the parameters of the posterior distribution using MCMC is of crucial importance,
but more so in light of the problem of degeneracy.
Here we have that, to a large extent, the exchange algorithm results in quite fast convergence
to the stationary distribution. We present a heuristic argument as to why this is the case. 
Assume that $s(\bfy) \in rint(C)$, that $\pi(\bftheta)$ is very flat and that $h(\cdot)$ is symmetric. 
We can then write the acceptance ratio (\ref{eqn:alpha}) as:
\begin{equation} 
\alpha\approx \min\left ( 1, \exp\left \{ (\bftheta - \bftheta')^t(s(\bfy')-s(\bfy)) \right \} \right )
\label{eqn:alpha2}
\end{equation}
where $s(\bfy)$ is the fixed vector of the sufficient statistics of the observed graph 
and $s(\bfy')$ is a vector of sufficient statistics of a graph drawn from $\pi(\bfy'|\bftheta')$.
It is known, see Section 2 of \cite{van:gil:han09}, for example, that  
\begin{equation} 
 (\bftheta -\bftheta')^t (\mu(\bftheta)-\mu(\bftheta')) \geq 0. 
\label{eqn:mean_para}
\end{equation}
Recall that $\mu(\bftheta)=\EE_{\bftheta}[s(\bfy)]$ is the mean parameterisation 
for $\bftheta$. 
Notice that if there is good agreement between $\bfy$ and $\bftheta'$, and good 
agreement between $\bfy'$ and $\bftheta$, then 
\[ s(\bfy') \approx \mu(\bftheta) \;\; \mbox{and}\;\;
   s(\bfy) \approx \mu(\bftheta').
\]
Equation (\ref{eqn:mean_para}) would then implies that the exponent in (\ref{eqn:alpha2}) is positive, 
whereby the exchange move is accepted. 
Empirically, it has been observed that fast convergence occurs even when the initial parameters are set 
to lie in the degenerate region. The right-hand plot of Figure~\ref{fig:2d-flo} displays the two dimensional 
trace of 12 MCMC runs starting
from very different points in the parameter space for the pedagogical example. In each instance, the Markov 
chain converges quickly to the high posterior density region. In fact, we have observed this behaviour for 
each of the datasets we analysed. 

\begin{figure}[htp]
\centering
\includegraphics[scale=0.6]{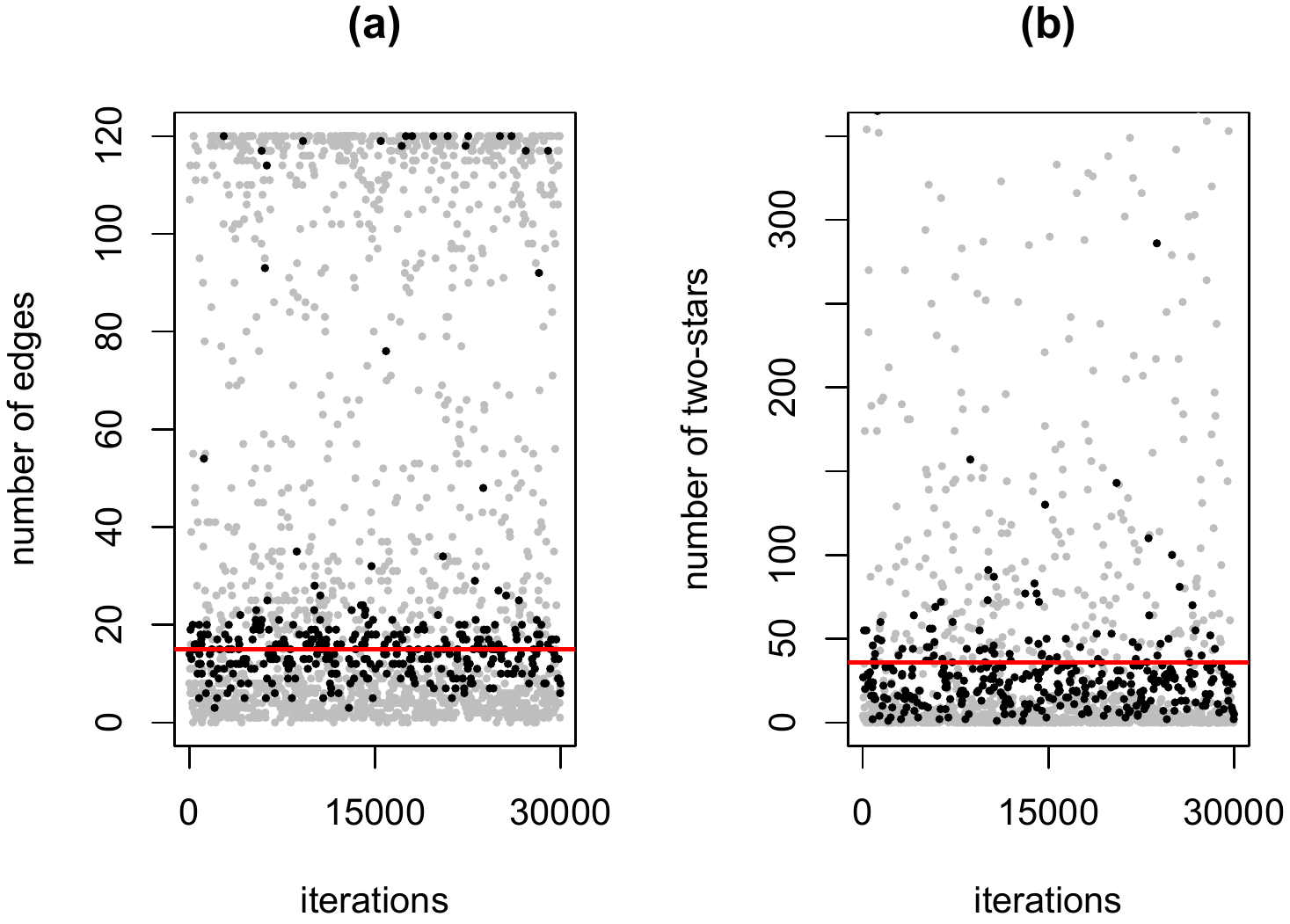}
\caption{Number of edges (left) and two-stars (right) of $2,000$ graphs (thinned by a factor of 15 
from the original $30,000$ iterations) simulated (gray dots) and graphs whose parameters 
were accepted (black dots). The solid line represents the number of edges in the observed graph.}
\label{fig:mixing-flo}
\end{figure}

Recall that the exchange algorithm involves drawing networks from the likelihood model.
It is natural to ask questions such as, what proportion of such simulated networks are full or empty?
And more importantly are any of these networks accepted in the MCMC algorithm?
Figure~\ref{fig:mixing-flo} partly answers these questions, in the case of the pedagogical example
of Section 3.4. Here it can be seen that the algorithm frequently simulates low or high density graphs,
but most of the networks accepted are those whose sufficient statistics are in close
agreement to the observed sufficient statistics. This is a very useful aspect of the algorithm.

\subsection{Population MCMC can improve mixing}
While easy to implement, the single-site update procedure can lead to 
slow mixing if there is strong temporal dependence in the state process.
As shown above, often there may be a strong correlation between model 
parameters and also the high posterior density region can be thin. 
In order to improve mixing we propose to use an adaptive direction sampling (ADS)
method (\cite{gil:rob:geo94} and \cite{rob:gil94}), similar 
to that of \cite{ter:vru08}.
We consider a population MCMC approach consisting of a collection of  $H$ chains which interact 
with one another.
Here the state space is $\{\bftheta_1,\dots,\bftheta_H\}$ with target distribution 
$\pi(\bftheta_1|\bfy) \otimes \cdots \otimes \pi(\bftheta_H|\bfy)$.
A ''parallel ADS'' move may be described algorithmically as follows.
\begin{itemize}
\item[ ] For each chain $h=1,\dots,H$
\begin{itemize}
\item[1.] Select at random $h_1$ and $h_2$ without replacement from $\{1,\dots,H\} \setminus h$
\item[2.] Sample $\bfepsilon$ from a symmetric distribution
\item[3.] Propose $\bftheta'_h=\bftheta^i_h + \gamma\left(\bftheta^i_{h_1}-\bftheta^i_{h_2}\right)+\bfepsilon$
\item[4.] Sample $\bfy'$ from $\pi(\cdot|\bftheta'_h)$ by MCMC methods, for example, the ``tie no tie'' (TNT) sampler, 
taking a realisation from a long run of the chain as an approximate draw from this distribution.
\item[5.] Accept the move from $\bftheta^i_h$ to $\bftheta^{i+1}_h = \bftheta'_h$ with probability
\begin{equation*}
\alpha=\min\left( 1, \frac{q_{\bftheta^i_h}(\bfy')\pi(\bftheta'_h) q_{\bftheta'_h}(\bfy)}
                          {q_{\bftheta^i_h}(\bfy)\pi(\bftheta^i_h) q_{\bftheta'_h}(\bfy')} 
           \right).
\end{equation*}
\end{itemize}
\end{itemize}

\begin{figure}[htp]
\centering
\includegraphics[scale=0.9]{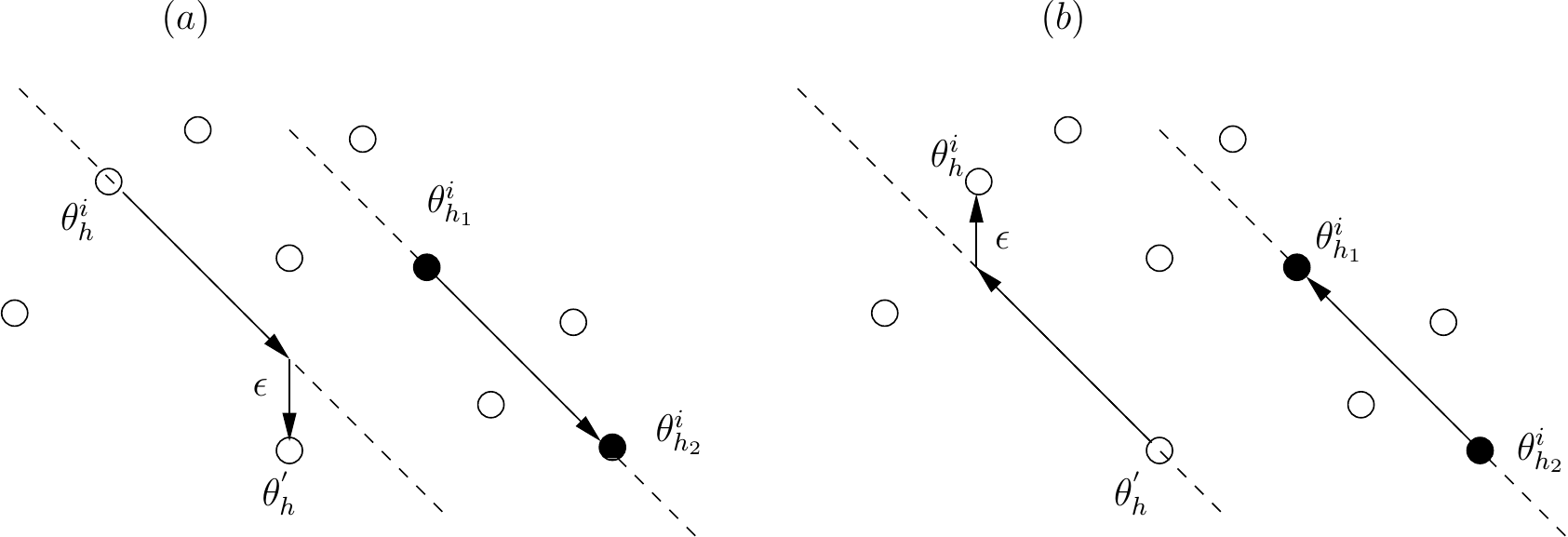}
\caption{(a) The parallel ADS move of $\bftheta^i_h$ is generated from
the difference of the states $\bftheta^i_{h_1}$ and $\bftheta^i_{h_2}$ selected from the other chains 
and a random term $\bfepsilon$. (b) The reverse jump is obtained by reversing 
$\epsilon$ and the order of the difference between $\bftheta^i_{h_1}$ and $\bftheta^i_{h_2}$.}
\label{fig:snooker}
\end{figure}

This move is illustrated graphically in Figure~\ref{fig:snooker}. 

For our running example we set $\gamma = 1$ and $\bfepsilon \sim \mathcal{N}(\mathbf{0},0.1\bfI_d)$
where the tuning parameters were set so that the acceptance rate is around $20\%$.
Initially all the chains are run in parallel using a block update of each individual chain and after a 
certain number of iterations the parallel ADS update begins. 
For this example we set the number of chains $H = 5$.
The posterior output from the population MCMC algorithm
is shown in Table~\ref{tab:popmcmc-flo}.

\begin{table}[htp]
\centering
\begin{tabular}{c|c|cc}
\multicolumn{4}{c}{} \\
\hline \hline
 & Parameter & Post. Mean & Post. Sd. \\ \hline 
\multirow{2}{*}{Chain 1} 
& $\theta_1$ (edges) & -2.39 & 0.55 \\
& $\theta_2$ (2-stars) & 0.11 & 0.12 \\ \hline
\multirow{2}{*}{Chain 2} 
& $\theta_1$ (edges) & -2.49 & 0.51 \\
& $\theta_2$ (2-stars) & 0.13 & 0.12 \\ \hline
\multirow{2}{*}{Chain 3} 
& $\theta_1$ (edges) & -2.43 & 0.57 \\
& $\theta_2$ (2-stars) & 0.11 & 0.12 \\ \hline 
\multirow{2}{*}{Chain 4} 
& $\theta_1$ (edges) & -2.44 & 0.54 \\
& $\theta_2$ (2-stars) & 0.13 & 0.13 \\ \hline 
\multirow{2}{*}{Chain 5} 
& $\theta_1$ (edges) & -2.42 & 0.52 \\
& $\theta_2$ (2-stars) & 0.12 & 0.13 \\ \hline 
\multirow{2}{*}{Overall} 
& $\theta_1$ (edges) & -2.44 & 0.54 \\
& $\theta_2$ (2-stars) & 0.12 & 0.12 \\ \hline \hline
\end{tabular}
\caption{Summary of posterior parameter density of the model (\ref{eqn:bay-two-star}).}
\label{tab:popmcmc-flo}
\end{table}

\begin{figure}[htp]
\centering
\includegraphics[scale=0.7]{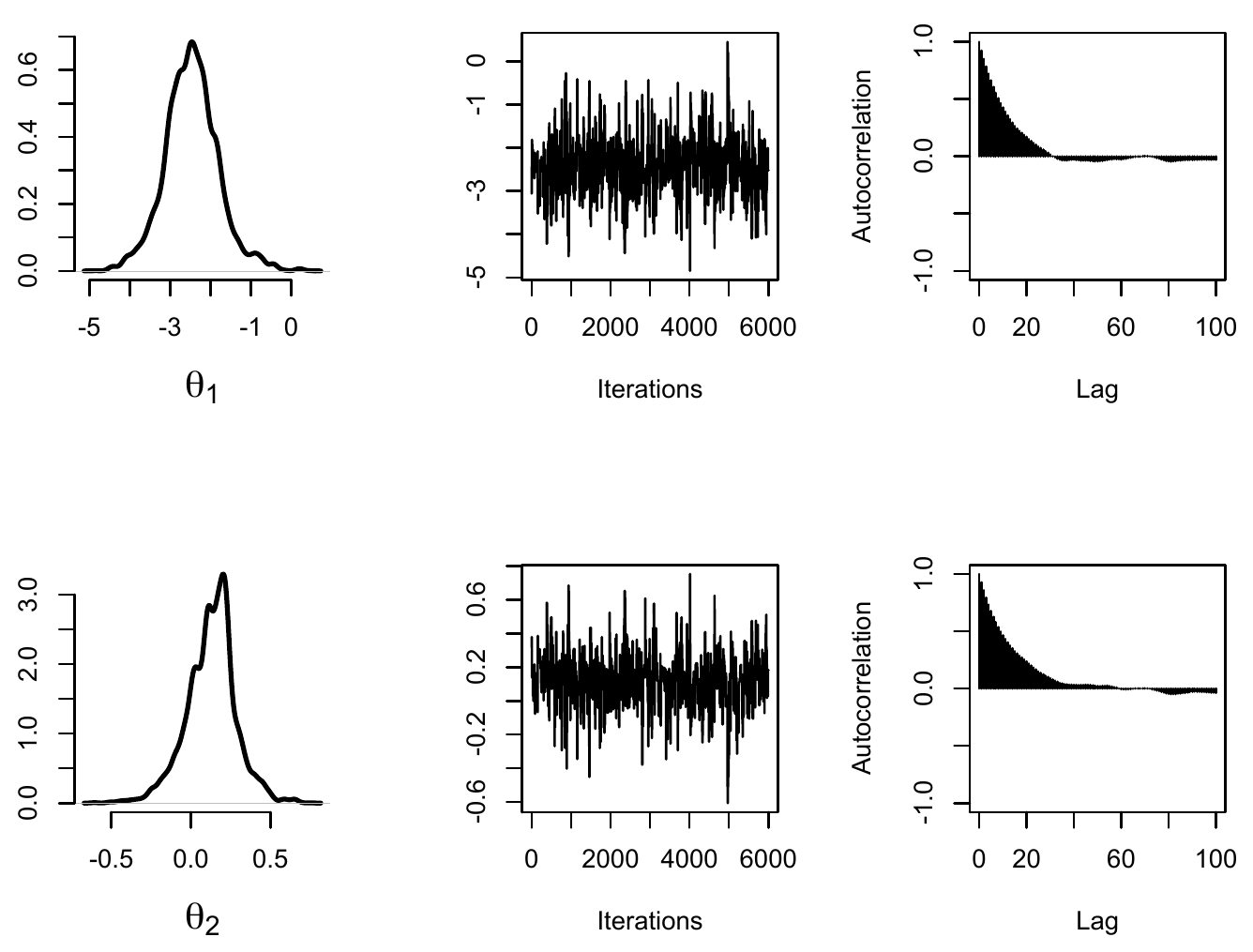}
\caption{MCMC output of a single chain of the model (\ref{eqn:bay-two-star}).}
\label{fig:popmcmc-flo}
\end{figure}

Figure~\ref{fig:popmcmc-flo} shows that the autocorrelation curve now decreases 
faster than the single-site update algorithm and is negligible at around lag 50. Recall that
for the single-site update version of this algorithm, the autocorrelation is negligible after
lag $150$. Therefore the single-site sampler would need to be run for around $3$ times more iterations 
than the population MCMC sampler in order to achieve a comparable number of effectively independent
draws from the posterior distribution. For more complex networks, the reduction in autocorrelation 
using the population MCMC algorithm was considerable. 

\begin{figure}[htp]
\centering
\includegraphics[scale=0.55]{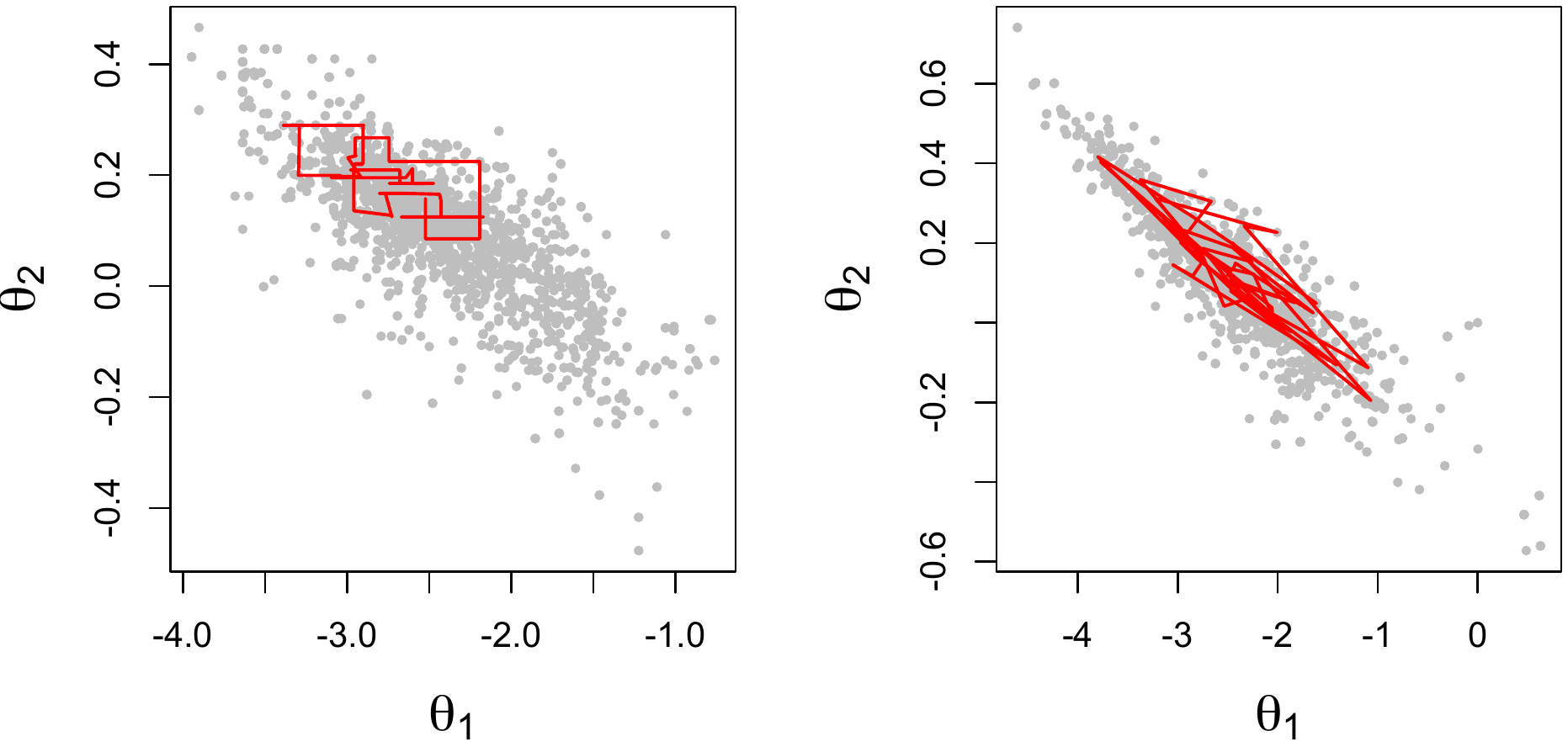}
\caption{2-D traces of 400 iterations for the single-site update algorithm (left) and a single chain of 
the parallel ADS update algorithm (right).}
\label{fig:ssuvspop-flo}
\end{figure}

Figure~\ref{fig:ssuvspop-flo} depicts the 2-dimensional trace of 400 iterations
for both the single-site update and  a  single chain using the parallel ADS update.
We see that the latter can easily move into the correlated high posterior density region
 thus ensuring a better mixing of the algorithm. 

The algorithm took approximately 13 minutes to estimate model (\ref{eqn:bay-two-star}) 
using $30,000$ iterations of the single-site update and about 6 minutes using the parallel ADS updater with 
$6,000$ iterations per chain.

\section{Examples}
We now consider three different benchmark networks datasets (two undirected and one directed) 
and we illustrate models with various specifications (see \cite{rob:pat:kal:lus07} 
and \cite{sni:pat:rob:han06} for details).
For each of them we apply the population MCMC with parallel ADS update, 
setting the number of chains equal to twice the number of model parameters.

\subsection{Molecule synthetic network}
\begin{figure}[htp]
\centering
\includegraphics[scale=0.7]{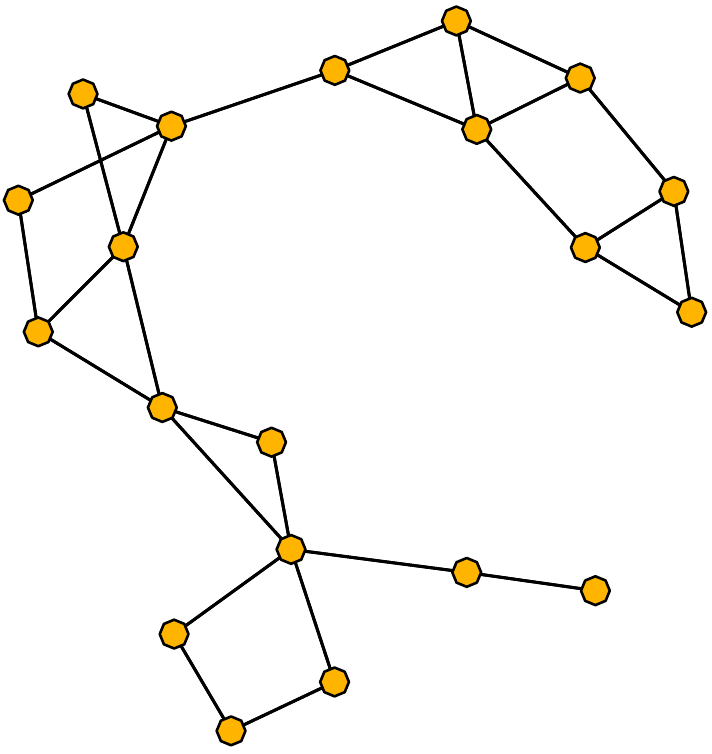}
\caption{Molecule synthetic graph.}
\label{fig:graph-mol}
\end{figure}

This dataset is included with the \texttt{ergm} package for \texttt{R}. 
The elongated shape of this 20-node graph, shown in Figure \ref{fig:graph-mol}, 
resembles the chemical structure of a molecule.

We consider the following 4-dimensional model 
\begin{equation}
\pi(\bftheta | \bfy) \propto \frac{1}{z(\bftheta)} 
\exp \left\lbrace \sum_{i=1}^{4} \theta_i s_{i}(\bfy) \right\rbrace \pi(\bftheta)
\label{eqn:bay-4-star}
\end{equation}
where
\begin{center}
\begin{tabular}{ll}
$s_{1}(\bfy) = \sum_{i<j}y_{ij}$ & number of edges\\
$s_{2}(\bfy) = \sum_{i<j<k}y_{ik}y_{jk}$ & number of two-stars\\
$s_{3}(\bfy) = \sum_{i<j<k<l}y_{il}y_{jl}y_{kl}$ & number of three-stars\\
$s_{4}(\bfy) = \sum_{i<j<k}y_{jk}y_{ik}y_{ij}$ & number of triangles
\end{tabular}
\end{center}
We use the flat prior $\pi( \theta_i ) \sim \mathcal{N}(0,30)$, for $i=1,\dots,4$, and we set 
$\gamma = 0.5$ and $\bfepsilon \sim \mathcal{N}(\mathbf{0},0.1\bfI)$ 
corresponding to an overall acceptance probability of $22\%$.
The auxiliary chain consists of $1,000$ iterations and the main one of $4,000$ iterations
for each chain.
The algorithm took approximately 9 minutes to sample the posterior model (\ref{eqn:bay-4-star})
for a total of $32,000$ iterations. 
In Table~\ref{tab:tabpopmcmc-mol} we see that the posterior density estimates of the 8 chains 
are similar to  each other.
Aggregating the output across all chains, the posterior estimates for the first three parameters 
indicate an overall tendency for nodes to 
have a limited number of multiple edges while the positive estimates relative to the triangle
parameter capture the propensity towards transitivity.
Table~\ref{tab:tabmcmle-mol} reports the estimates of both MC-MLE and MPLE. 
Here MC-MLE fails to converge as the MPLE estimates generate mainly full graphs.

\begin{figure}[htp]
\centering
\includegraphics[scale=0.7]{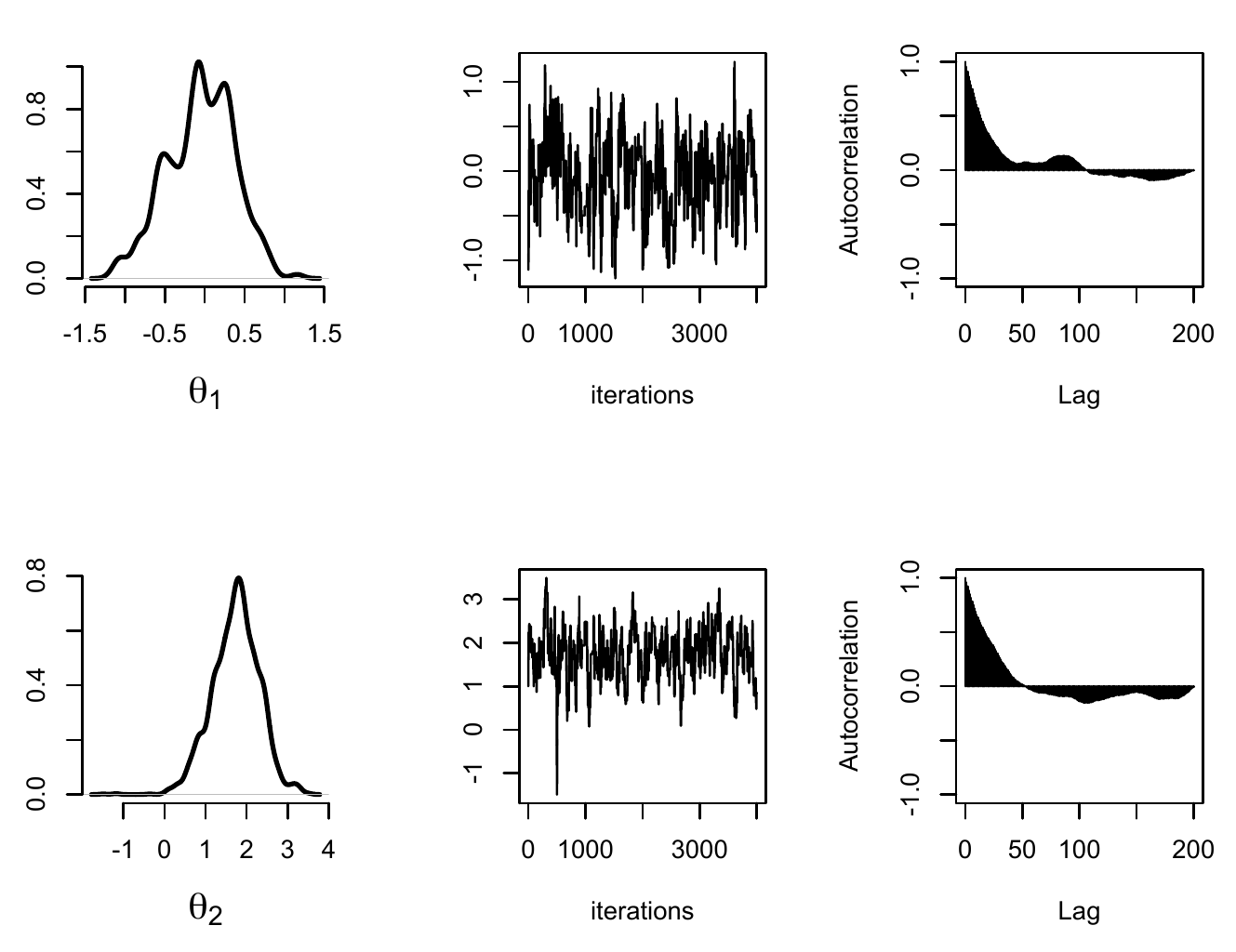}
\\
\vspace{.3cm}
\includegraphics[scale=0.7]{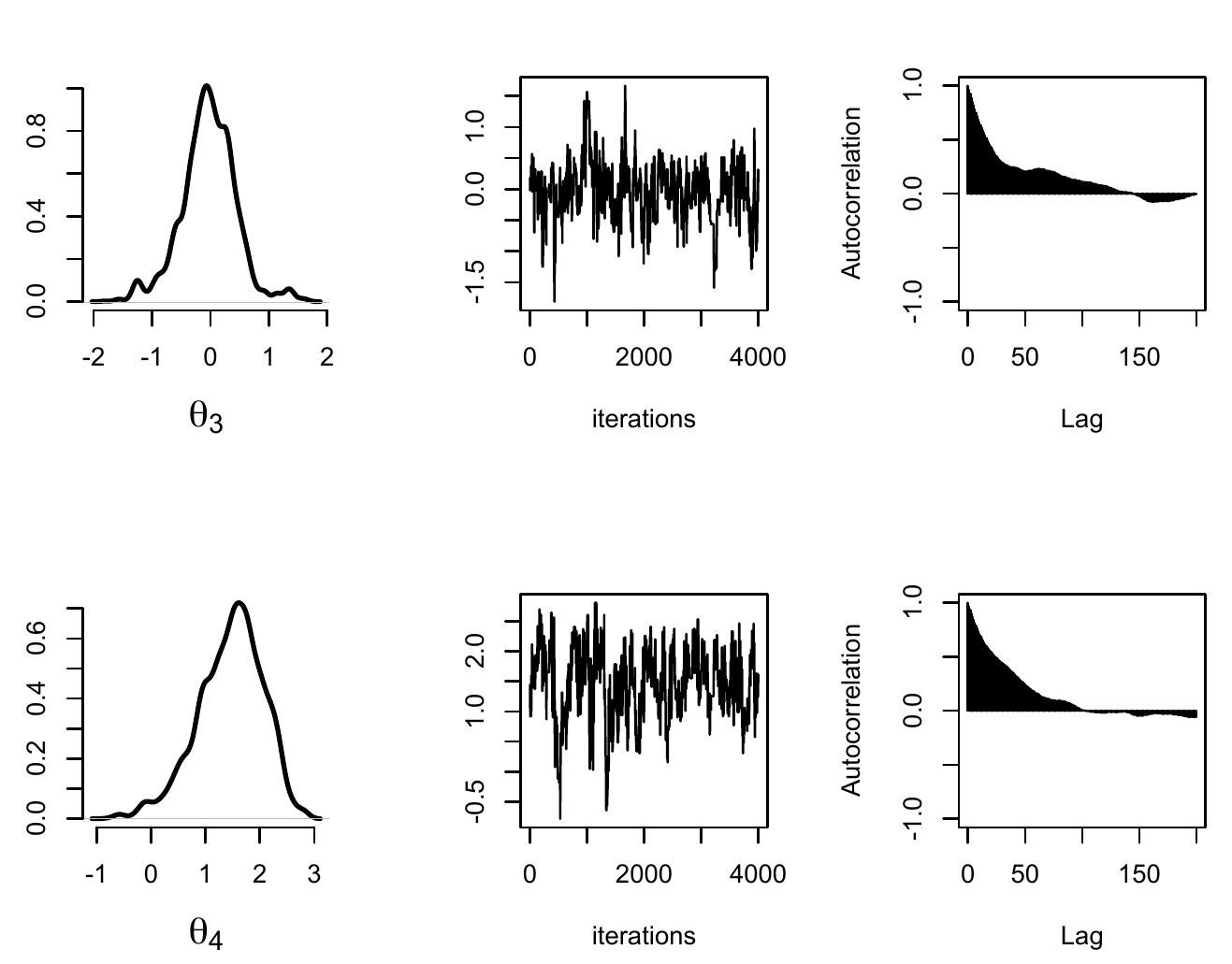}
\caption{Molecule dataset: MCMC output of a single chain of the posterior distribution (\ref{eqn:bay-4-star}).}
\label{fig:popmcmc-mol}
\end{figure}

In Figure \ref{fig:popmcmc-mol} it can be seen that the autocorrelations 
of the parameters decay quickly around lag $200$. By comparison, a single chain sampler (not reported here) 
with single-site updating of the $\bftheta$ vector led to negligible autocorrelation at around lag $1,000$. 
Therefore, the single-site sampler would need to be run for around $5$ times more iterations 
than the population MCMC sampler in order to achieve comparable number of effectively independent
draws from the posterior distribution.

\begin{table}[htp]
\centering
\begin{tabular}{c|c|cc}
\multicolumn{4}{c}{} \\
\hline \hline
 & Parameter & Post. Mean & Post. Sd. \\ \hline 
\multirow{4}{*}{Chain 1} 
& $\theta_1$ (edges) & 2.88 & 3.10 \\
& $\theta_2$ (2-stars) & -1.06 & 0.92 \\ 
& $\theta_3$ (3-stars) & -0.04 & 0.43 \\ 
& $\theta_4$ (triangles) & 1.55 & 0.54 \\ 
\hline 
\multirow{4}{*}{Chain 2} 
& $\theta_1$ (edges) & 2.61 & 3.12 \\
& $\theta_2$ (2-stars) & -1.02 & 0.99 \\ 
& $\theta_3$ (3-stars) & -0.05 & 0.42 \\ 
& $\theta_4$ (triangles) & 1.70 & 0.57 \\ 
\hline 
\multirow{4}{*}{Chain 3} 
& $\theta_1$ (edges) & 2.71 & 3.13 \\
& $\theta_2$ (2-stars) & -1.01 & 0.96 \\ 
& $\theta_3$ (3-stars) & -0.06 & 0.48 \\ 
& $\theta_4$ (triangles) & 1.58 & 0.53 \\ 
\hline 
\multirow{4}{*}{Chain 4} 
& $\theta_1$ (edges) & 2.68 & 3.35 \\
& $\theta_2$ (2-stars) & -1.01 & 1.02 \\ 
& $\theta_3$ (3-stars) & -0.03 & 0.47 \\ 
& $\theta_4$ (triangles) & 1.45 & 0.59 \\ 
\hline 
\multirow{4}{*}{Chain 5} 
& $\theta_1$ (edges) & 2.72 & 3.11 \\
& $\theta_2$ (2-stars) & -1.10 & 1.14 \\ 
& $\theta_3$ (3-stars) & -0.02 & 0.44 \\ 
& $\theta_4$ (triangles) & 1.59 & 0.63 \\ 
\hline 
\multirow{4}{*}{Chain 6} 
& $\theta_1$ (edges) & 2.87 & 3.46 \\
& $\theta_2$ (2-stars) & -1.05 & 1.00 \\ 
& $\theta_3$ (3-stars) & -0.10 & 0.47 \\ 
& $\theta_4$ (triangles) & 1.65 & 0.55 \\ 
\hline 
\multirow{4}{*}{Chain 7} 
& $\theta_1$ (edges) & 2.60 & 3.25 \\
& $\theta_2$ (2-stars) & -1.10 & 1.06 \\ 
& $\theta_3$ (3-stars) & -0.10 & 0.47 \\ 
& $\theta_4$ (triangles) & 1.62 & 0.51 \\ 
\hline 
\multirow{4}{*}{Chain 8} 
& $\theta_1$ (edges) & 2.82 & 3.64 \\
& $\theta_2$ (2-stars) & -1.03 & 1.03 \\ 
& $\theta_3$ (3-stars) & -0.10 & 0.50 \\ 
& $\theta_4$ (triangles) & 1.61 & 0.59 \\ 
\hline 
\multirow{4}{*}{Overall} 
& $\theta_1$ (edges) & 2.72 & 3.27 \\
& $\theta_2$ (2-stars) & -1.02 & 1.02 \\ 
& $\theta_3$ (3-stars) & -0.05 & 0.46 \\ 
& $\theta_4$ (triangles) & 1.60 & 0.57 \\ 
\hline \hline
\end{tabular}
\caption{Molecule dataset: summary of posterior parameter density of the molecule network (\ref{eqn:bay-4-star}).}
\label{tab:tabpopmcmc-mol}
\end{table}


\begin{table}[htp]
\centering
\begin{tabular}{c|cc|cc}
 \multicolumn{1}{c}{} & \multicolumn{2}{c}{MC-MLE} & \multicolumn{2}{c}{MPLE}\\
\hline \hline
Parameter & Estimate & Std. Error & Estimate & Std. Error \\ \hline
$\theta_1$ (edges) & 4.11 & NA & 5.08 & 1.90\\
$\theta_2$ (2-stars) & -1.64 & NA & -2.02 & 0.60\\
$\theta_3$ (3-stars) & 0.42 & NA & 0.52 & 0.27\\
$\theta_4$ (triangles) & 1.30 & NA & 1.60 & 0.39\\
\hline \hline
\end{tabular}
\caption{Molecule dataset: Summary of parameter estimates of the model (\ref{eqn:bay-4-star}).}
\label{tab:tabmcmle-mol}
\end{table}

\begin{figure}[htp]
\centering
\includegraphics[scale=0.7]{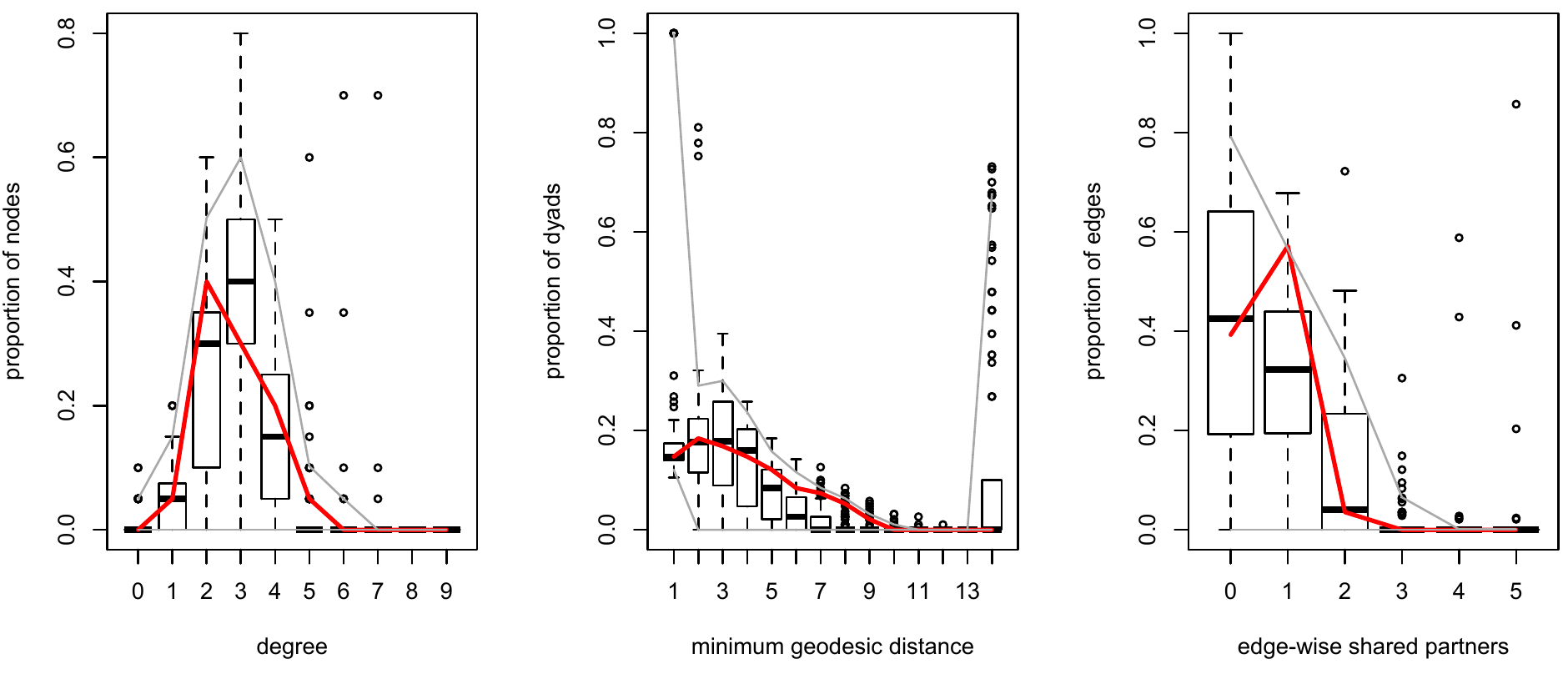}
\caption{Molecule dataset: Bayesian goodness-of-fit output.}
\label{fig:bgof-mol}
\end{figure}

A pragmatic way to examine the fit of the data to the posterior model the output obtained is to implement a 
Bayesian goodness-of-fit procedure. In order to do this, 100 graphs are simulated from 100 independent 
realisations taken from the estimated posterior distribution and compared to the observed
graph in terms of high-level characteristics which are not modelled explicitly, namely, degree distributions
(for degrees greater than 3); the minimum geodesic distance (namely, the proportion of pairs of nodes whose shortest 
connected path is of length $l$, for $l=1,2,\dots$); the number of edge-wise shared partners 
(the number of edges in the network that share $l$ neighbours in common for $l=1,2,\dots$.)
The results are shown in Figure \ref{fig:bgof-mol} and indicate that the structure of the 
observed graph can be considered as a possible realisation of the posterior density.

\subsection{Dolphins network}

\begin{figure}[htp]
\centering
\includegraphics[scale=0.55]{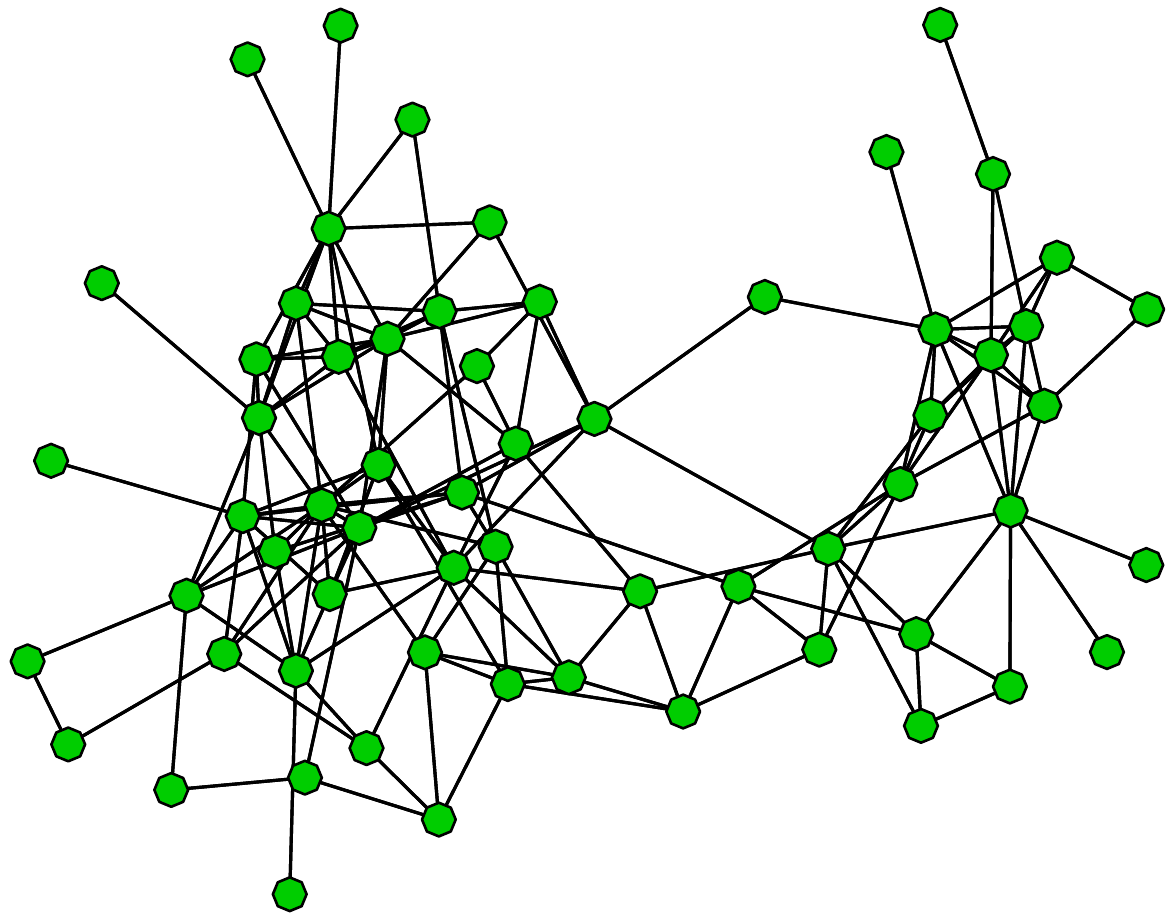}
\caption{Dolphins network dataset.}
\label{fig:graph-dol}
\end{figure}

The undirected graph displayed in Figure~\ref{fig:graph-dol} represents social associations 
between 62 dolphins living off Doubtfull Sound in New Zealand \citep{lus:sch:boi:haa:slo:daw03}.
The graph is inhomogeneous, a few nodes have large number of edges and many have only one or two edges.

We propose to estimate the following 3-dimensional model using three new specifications
proposed by \cite{sni:pat:rob:han06}: 
\begin{equation}
\pi(\bftheta | \bfy) \propto \frac{1}{z(\bftheta)} 
\exp \left\lbrace 
\theta_1 s_{1}(\bfy) + 
\theta_2 u(\bfy,\phi_u) + 
\theta_3 v(\bfy,\phi_v) 
\right\rbrace  
\pi(\bftheta)
\label{eqn:dol-newspec}
\end{equation}
where
\begin{center}
\begin{tabular}{ll}
$s_{1}(\bfy) = \sum_{i<j}y_{ij}$ & number of edges\\
$u(\bfy,\phi_u) = 
e^{\phi_u} \sum_{i=1}^{n-1} 
\left\{ 1- \left( 1 - e^{-\phi_u} \right )^{i} \right \} D_i(\bfy)$ 
& geometrically weighted degree\\
$v(\bfy,\phi_v) = 
e^{\phi_v} \sum_{i=1}^{n-2}
\left \{ 1-\left( 1 - e^{-\phi_v} \right)^{i} \right \} EP_i(\bfy)$ 
& geometrically weighted edgewise\\
& shared partner.
\end{tabular}
\end{center}
We set $\phi_{u}=0.8$ and $\phi_{v}=0.8$ so that the model is a regular, that is, a non-curved 
exponential random graph model \citep{hun:han06}.
We use the flat multivariate normal prior $\pi( \bftheta ) \sim \mathcal{N}(\mathbf{0},30 \bfI)$ and 
we set $\gamma = 0.5$ and $\bfepsilon \sim \mathcal{N}(\mathbf{0},0.1\bfI)$, where
the tuning parameters were chosen so that the overall acceptance rate was around $10\%$. 
The auxiliary chain consists of $15,000$ iterations and the main chain $10,000$ iterations for each of the
6 chains of the population.
The algorithm took approximately 1 hour and 10 minutes to estimate model (\ref{eqn:dol-newspec}).
A single-site algorithm would take many more hours to carry out estimation using the same 
overall number of iterations.
From Table~\ref{tab:tabpopmcmc-dol} it can seen that mixing across the chains appears to be
reasonable and the overall posterior estimates displayed in the lower block of 
Table~\ref{tab:tabpopmcmc-dol} provide us with a useful interpretation of the observed graph.
The tendency to a low density of edges expressed by the negative posterior mean of the
first parameter, is balanced by a propensity to multiple local clustering and tied nodes to share
multiple neighbours in common expressed, respectively, by the last two positive posterior parameter 
estimates.

\begin{table}[htp]
\centering
\begin{tabular}{c|c|cc}
\multicolumn{4}{c}{} \\
\hline \hline
 & Parameter & Post. Mean & Post. Sd. \\ \hline 
\multirow{3}{*}{Chain 1} 
& $\theta_1$ (edges) &-4.24 & 0.32 \\
& $\theta_2$ (gwd) & 1.28 & 0.50 \\ 
& $\theta_3$ (gwesp) & 0.94 & 0.13 \\ 
\hline 
\multirow{3}{*}{Chain 2} 
& $\theta_1$ (edges) &-4.24 & 0.34 \\
& $\theta_2$ (gwd) & 1.27 & 0.50 \\ 
& $\theta_3$ (gwesp) & 0.94 & 0.13 \\ 
\hline 
\multirow{3}{*}{Chain 3} 
& $\theta_1$ (edges) &-4.27 & 0.33 \\
& $\theta_2$ (gwd) & 1.32 & 0.50 \\ 
& $\theta_3$ (gwesp) & 0.95 & 0.13 \\   
\hline 
\multirow{3}{*}{Chain 4} 
& $\theta_1$ (edges) &-4.27 & 0.35 \\
& $\theta_2$ (gwd) & 1.27 & 0.52 \\ 
& $\theta_3$ (gwesp) & 0.95 & 0.13 \\ 
\hline 
\multirow{3}{*}{Chain 5} 
& $\theta_1$ (edges) &-4.30 & 0.37 \\
& $\theta_2$ (gwd) & 1.32 & 0.55 \\ 
& $\theta_3$ (gwesp) & 0.95 & 0.13 \\ 
\hline 
\multirow{3}{*}{Chain 6} 
& $\theta_1$ (edges) &-4.24 & 0.34 \\
& $\theta_2$ (gwd) & 1.32 & 0.51 \\ 
& $\theta_3$ (gwesp) & 0.95 & 0.13 \\ 
\hline 
\multirow{3}{*}{Overall} 
& $\theta_1$ (edges) &-4.27 & 0.35 \\
& $\theta_2$ (gwd) & 1.30 & 0.52 \\ 
& $\theta_3$ (gwesp) & 0.95 & 0.13 \\ 
\hline \hline
\end{tabular}
\caption{Dolphins dataset: summary of posterior parameter density of the model (\ref{eqn:dol-newspec}).}
\label{tab:tabpopmcmc-dol}
\end{table}

\begin{figure}[htp]
\centering
\includegraphics[scale=0.7]{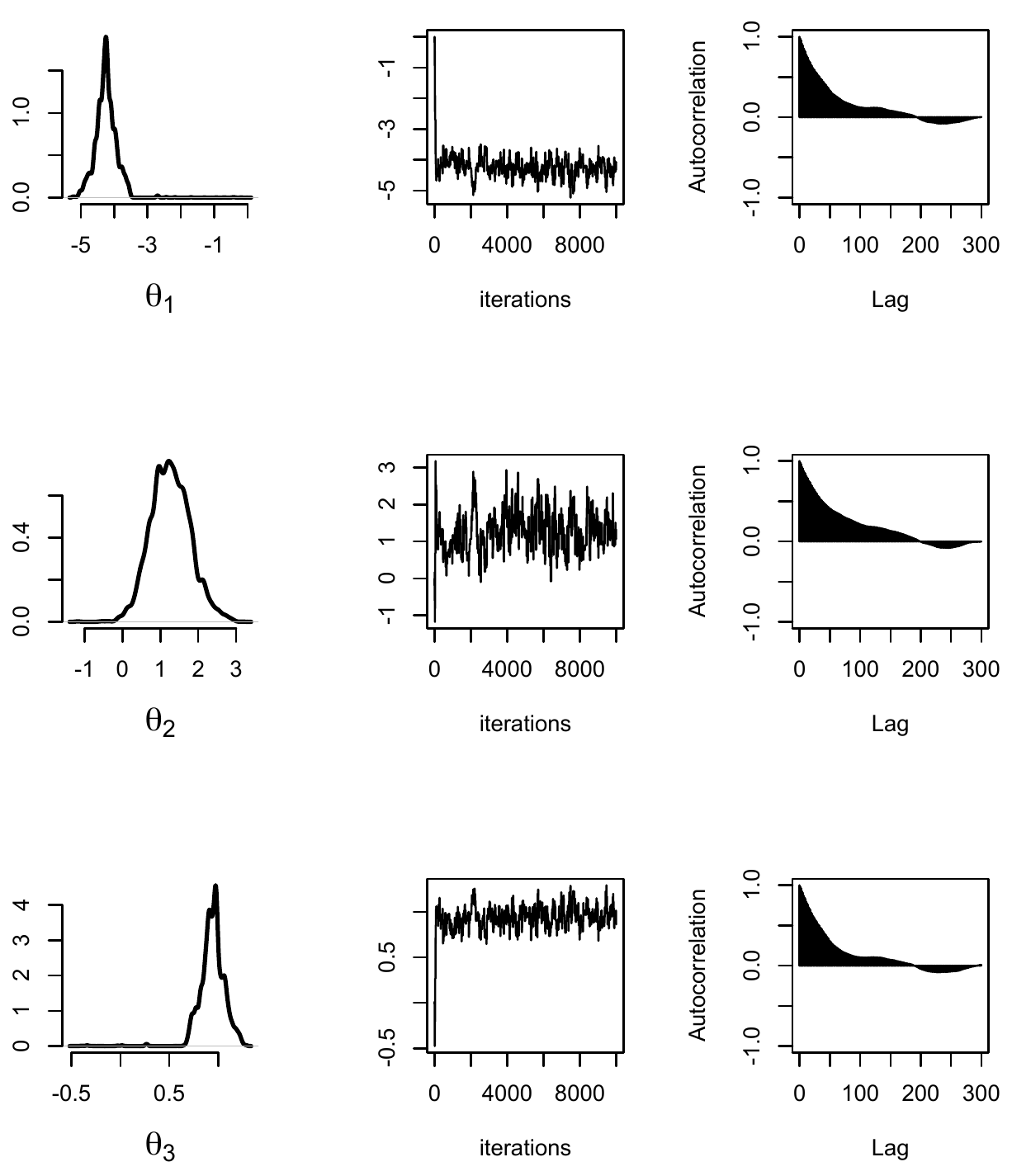}
\caption{Dolphins dataset: MCMC output of a single chain of the model (\ref{eqn:dol-newspec}).}
\label{fig:popmcmc-dol}
\end{figure}

Posterior estimates and autocorrelation function plots are reported in Figure~\ref{fig:popmcmc-dol}.
Each of the 3 autocorrelation functions decrease quite quickly and are negligible at around lag 200. 
A single chain MCMC version of the algorithm (not reported here) led to autocorrelation functions which 
decayed after lag $1500$ -- roughly a $7$ fold increases with respect to the population MCMC version 
of the algorithm.  

As an aside, note that both the MC-MLE and MPLE estimates were quite similar to the posterior mean
estimates for this dataset. 

\begin{figure}[htp]
\centering
\includegraphics[scale=0.55]{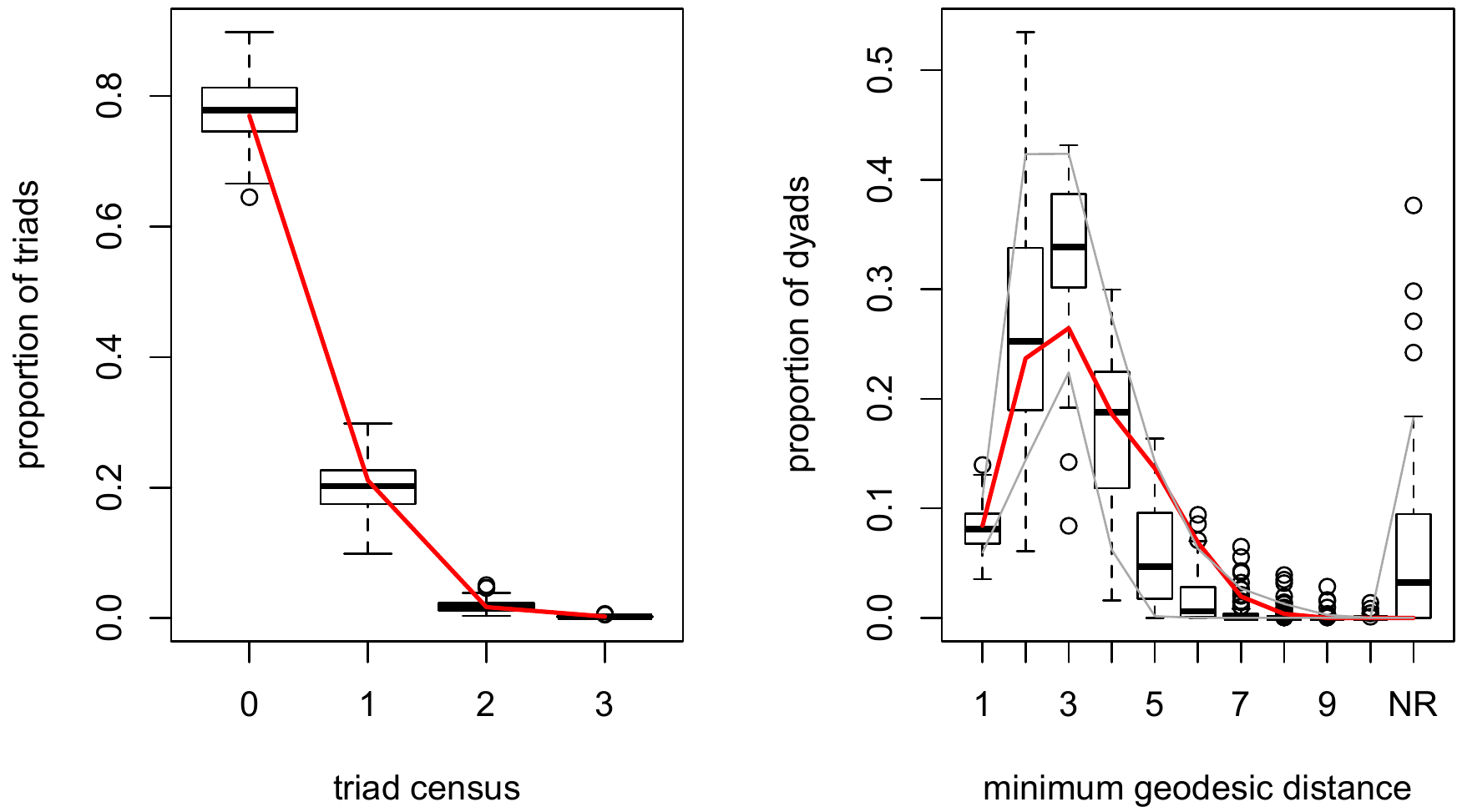}
\caption{Dolphins dataset: Bayesian goodness-of-fit output.}
\label{fig:bgof-dol}
\end{figure}

Similar to the previous example, we carry out a Bayesian goodness of fit test. The results of this are
displayed in Figure~\ref{fig:bgof-dol} and suggest that the model is a reasonable fit to the data. 

\subsection{Sampson's Monk network}

\begin{figure}[htp]
\centering
\includegraphics[scale=0.55]{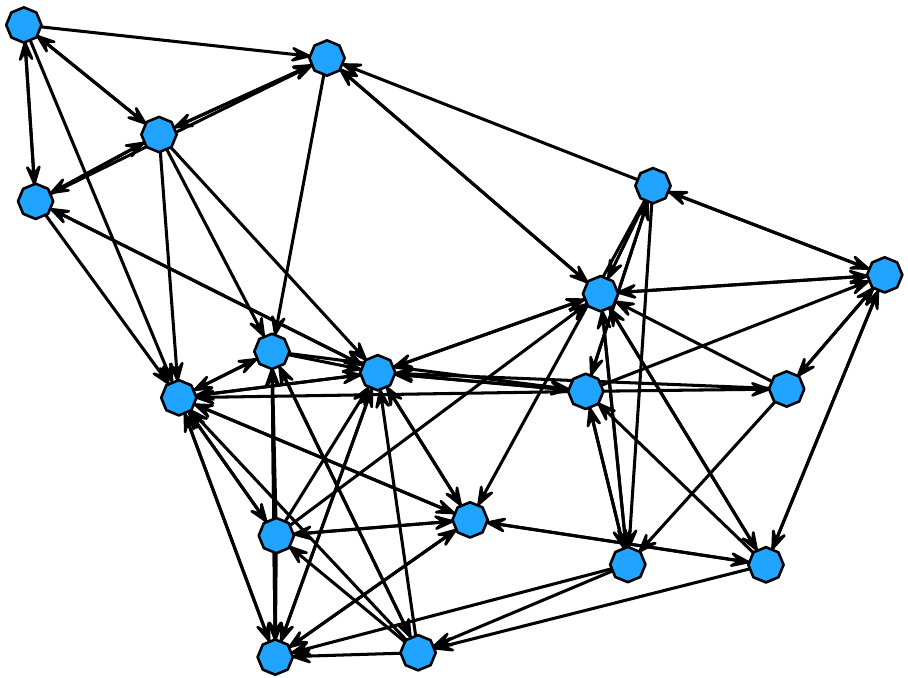}
\caption{Sampson's Monk network.}
\label{fig:graph-mon}
\end{figure}

This network represents the liking relationships among a group of 18 novices 
who were preparing to join a monastic order \citep{sam68}. This network consists
of directed edges between actors, and is presented in Figure~\ref{fig:graph-mon}.

We propose to estimate the following 3-dimensional model
\begin{equation}
\pi(\bftheta | \bfy) \propto \frac{1}{z(\bftheta)} 
\exp \left\lbrace \theta_1 s_{1}(\bfy) + \theta_2 s_{2}(\bfy) + \theta_3  s_{3}(\bfy)
\right\rbrace \pi(\bftheta)
\label{eqn:sam-ctriple}
\end{equation}
where
\begin{center}
\begin{tabular}{ll}
$s_{1}(\bfy) = \sum_{i<j}y_{ij}$ & number of edges\\
$s_{2}(\bfy) = \sum_{i<j}y_{ij}y_{ji}$ & number of mutual edges\\
$s_{3}(\bfy) = \sum_{i<j}y_{ij}y_{jk}y_{ki}$ & number of cyclic triads.
\end{tabular}
\end{center}
We use the flat multivariate normal prior $\pi( \bftheta ) \sim \mathcal{N}(\mathbf{0},30 \bfI_3 )$ 
and we set $\gamma = 0.8$ 
and $\bfepsilon \sim \mathcal{N}(\mathbf{0},0.1\bfI_3)$, giving an overall acceptance rate of $18\%$. 
The auxiliary chain consists of $2,000$ iterations and the main one of $5,000$ iterations for each 
of the 6 chains of the population.
Notice in Table~\ref{tab:tabpopmcmc-sam} that there is very little difference in output between the
chains, giving good evidence that there is reasonable mixing between parameters across the chains.
Table~\ref{tab:tabpopmcmc-sam} shows that the posterior estimate for the cyclic triads parameter
is not significative. The tendency to a low number of edges as expressed by the
edge parameter is balanced by a high level of reciprocity expressed by mutual parameter.

\begin{table}[htp]
\centering
\begin{tabular}{c|c|cc}
\multicolumn{4}{c}{} \\
\hline \hline
 & Parameter & Post. Mean & Post. Sd. \\ \hline 
\multirow{3}{*}{Chain 1} 
& $\theta_1$ (edges) &-1.72 & 0.31 \\
& $\theta_2$ (mutual) & 2.34 & 0.43 \\ 
& $\theta_3$ (ctriple) &-0.05 & 0.16 \\ 
\hline 
\multirow{3}{*}{Chain 2} 
& $\theta_1$ (edges) &-1.73 & 0.30 \\
& $\theta_2$ (mutual) & 2.38 & 0.42 \\ 
& $\theta_3$ (ctriple) &-0.02 & 0.16 \\ 
\hline 
\multirow{3}{*}{Chain 3} 
& $\theta_1$ (edges) &-1.74 & 0.29 \\
& $\theta_2$ (mutual) & 2.37 & 0.43 \\ 
& $\theta_3$ (ctriple) &-0.04 & 0.15 \\ 
\hline 
\multirow{3}{*}{Chain 4} 
& $\theta_1$ (edges) &-1.72 & 0.29 \\
& $\theta_2$ (mutual) & 2.33 & 0.44 \\ 
& $\theta_3$ (ctriple) &-0.06 & 0.16 \\ 
\hline 
\multirow{3}{*}{Chain 5} 
& $\theta_1$ (edges) &-1.73 & 0.30 \\
& $\theta_2$ (mutual) & 2.30 & 0.43 \\ 
& $\theta_3$ (ctriple) &-0.06 & 0.16 \\ 
\hline 
\multirow{3}{*}{Chain 6} 
& $\theta_1$ (edges) &-1.72 & 0.30 \\
& $\theta_2$ (mutual) & 2.27 & 0.44 \\ 
& $\theta_3$ (ctriple) &-0.06 & 0.16 \\ 
\hline 
\multirow{3}{*}{Overall} 
& $\theta_1$ (edges) &-1.72 & 0.30 \\
& $\theta_2$ (mutual) & 2.33 & 0.43 \\ 
& $\theta_3$ (ctriple) &-0.04 & 0.16 \\ 
\hline \hline
\end{tabular}
\caption{Monks dataset: Summary of posterior parameter density of the model (\ref{eqn:sam-ctriple}).}
\label{tab:tabpopmcmc-sam}
\end{table}

\begin{figure}[htp]
\centering
\includegraphics[scale=0.7]{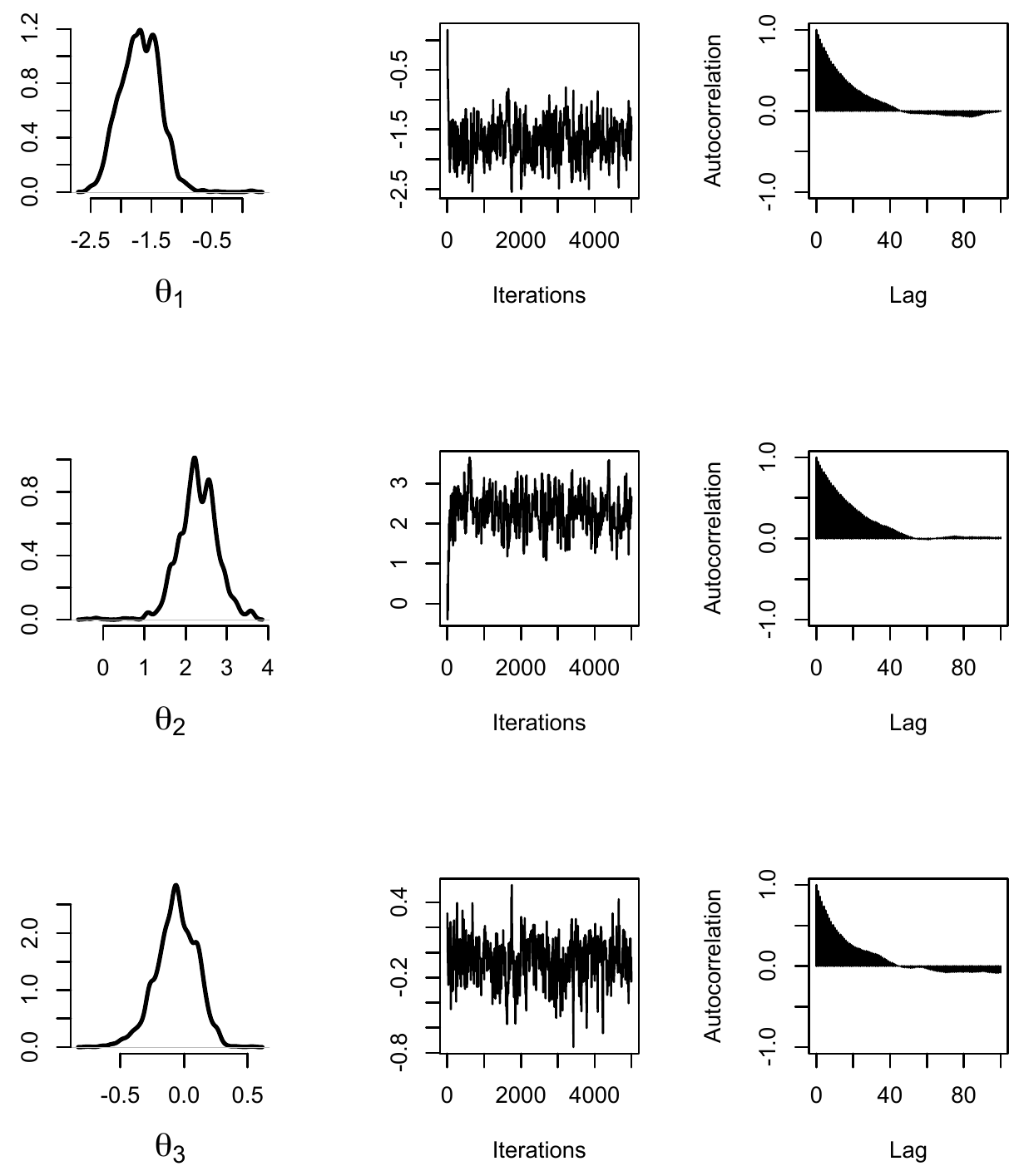}
\caption{Monks dataset: MCMC output of a single chain from (\ref{eqn:sam-ctriple}).}
\label{fig:popmcmc-sam}
\end{figure}

Figure~\ref{fig:popmcmc-sam} displays posterior density estimates from a single chain of the 
MCMC algorithm, together with autocorrelation functions for each parameter. All 3 autocorrelation 
functions decrease quite quickly and are negligible at around lag 60. This behaviour was very similar
to each of the other 5 chains. By comparison, a single chain MCMC run (not reported here) with an equivalent number of
iterations led to significantly higher autocorrelations, which were negligible at lag $400$.
Roughly $6$ or $7$ times more iterations for a single chain run would therefore be needed to yield effectively
the same number of independent draws from the posterior as the population MCMC version of the algorithm.

Here the algorithm took approximately 8 minutes to estimate model (\ref{eqn:sam-ctriple}).
As for the previous example, we note that the MC-MLE and MPLE estimates are similar to the
posterior mean estimates. 

\begin{figure}[htp]
\centering
\includegraphics[scale=0.7]{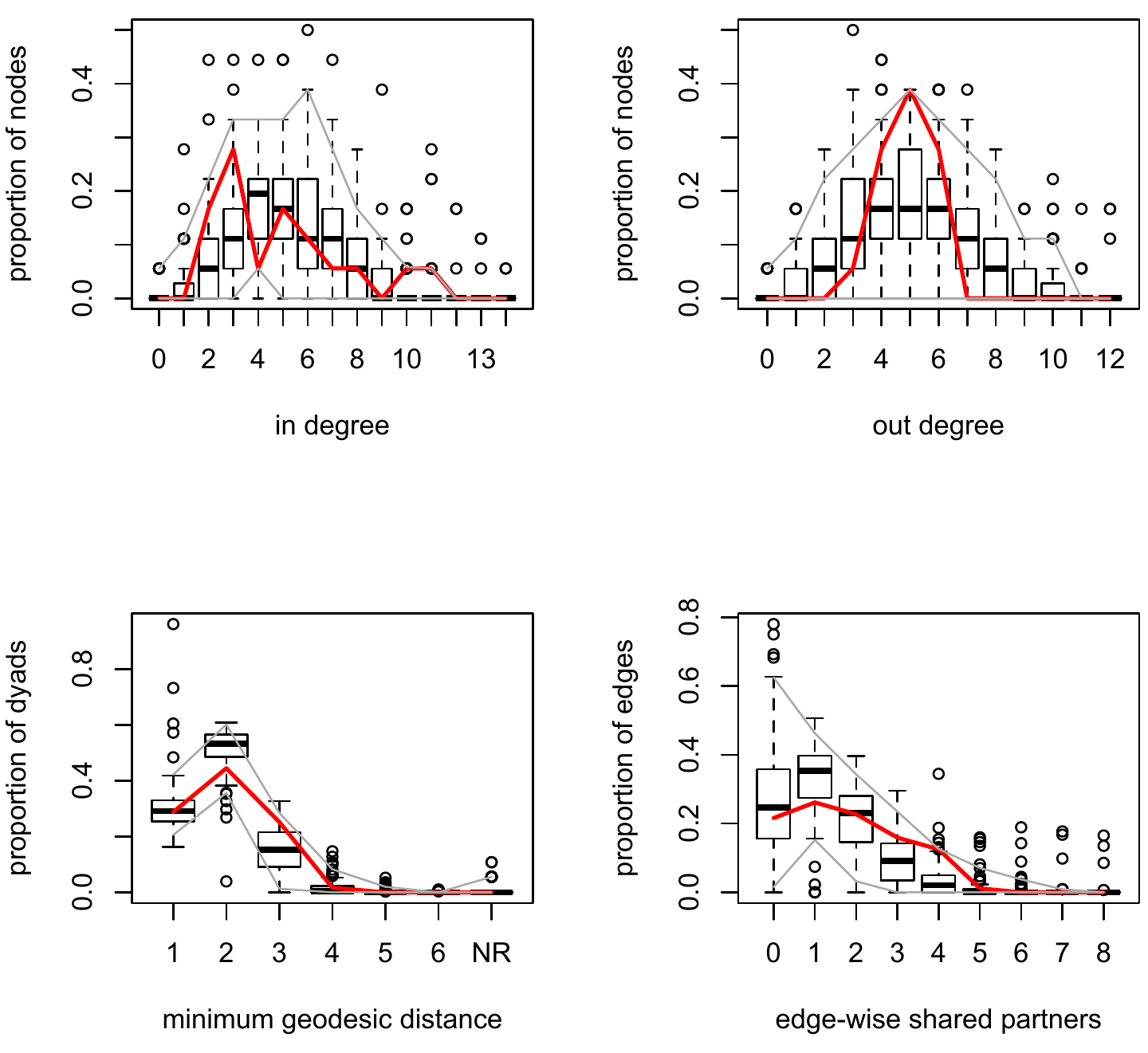}
\caption{Monks dataset: Bayesian goodness-of-fit output.}
\label{fig:bgof-sam}
\end{figure}

As before, we propose a series of Bayesian goodness-of-fit tests to understand how well 
the estimated model fits a set of observations. The results in Figure~\ref{fig:bgof-sam}
suggest that the model is a reasonable fit to the data. Since the network has
directed edges, we used in-degree and out-degree statistics, instead of degree distributions. 

\section{Discussion}
\label{sec:conc}

This paper has presented a Bayesian inferential framework for social network analysis for exponential
random graph models. 
The approach used here, based on the exchange algorithm \cite{Murray06}, is shown to give very good performance.
Here we present some datasets for which classical inference approaches, Maximum pseudolikelihood estimation and
Monte Carlo - Maximum likelihood estimation, both of which are standard practice in social network analysis,
fail to give reasonable parameter estimates. By contrast the Bayesian approach yield parameter estimates
consistent with the data, in the sense that networks simulated from the posterior had similar topologies
to the observed data. 
 
The issue of model degeneracy is important for ERGMs, and one which can cause potential difficulties
for inference methods, especially for the MC-MLE method -- a poorly chosen set of initial parameters,
from the degeneracy region will lead to poor estimation. Empirical evidence presented in this paper, 
suggests that the method we propose here results in a Markov chain which, even if initialised with
parameters from the degenerate region, will quickly converge to the posterior high density regions.

Our analysis has shown that the high posterior density region of ERGMs has typically very thin and 
correlated
support. This point is well addressed in \cite{rin:fie:zho09}, for example. This presents a very difficult
situation for exploration of the posterior support using MCMC methods. The population MCMC method
presented in this paper is a useful first step towards addressing this problem, and further
exploration in this direction should prove useful. All of the methods used in this paper are
presented in the \texttt{R} package \texttt{Bergm} which accompanies this paper, and should prove
useful to practitioners.

Estimation for larger networks is feasible but at the cost of increased computational time. In experiments 
not reported here, we carried out inference for a graph with $104$ nodes in under $2$ hours. 
Generally speaking, it seems reasonable to expect that the number of iterations needed for the auxiliary chain 
should be proportional to the number of dyads of the graphs, $n^2$ for a graph with $n$ nodes. This is
the computational bottleneck of the algorithm. The population MCMC approach which we outlined is very well
suited to parallel computing, and this may in turn lead to reduction in the computational time needed to
service ERGMs using the algorithm described in this paper.

\paragraph{Acknowledgement} 
Alberto Ciamo was supported by an IRCSET Embark Initiative award and Nial Friel's
research was supported by a Science Foundation Ireland Research Frontiers Program grant, 09/RFP/MTH2199. 
The authors wish to acknowledge Johan Koskinen and Michael Schwienberger  
for helpful comments on an earlier draft of this paper and also to thank the 
participants of the workshop \textit{Statistical Methods for the Analysis of Network 
Data in Practice} in Dublin, June 2009, for useful feedback and constructive comments 
on this work. 
\bibliographystyle{asa} 
\bibliography{myref}
\end{document}